\documentclass[10pt,twocolumn,letterpaper]{article}

\usepackage{cvpr}
\usepackage{times}
\usepackage{epsfig}
\usepackage{graphicx}
\usepackage{amsmath}
\usepackage{amssymb}
\usepackage{subfigure}
\usepackage{makecell}

\usepackage{multirow}

\usepackage{threeparttable}
\usepackage{kotex}
\usepackage{xcolor}
\usepackage{footnote}

\usepackage{arydshln}
\usepackage{balance}

\definecolor{cadmiumgreen}{rgb}{0.0, 0.42, 0.24}
\definecolor{cadmiumred}{rgb}{0.89, 0.0, 0.13}
\newcommand{\green}[1]{\textcolor{cadmiumgreen}{#1}}
\newcommand{\red}[1]{\textcolor{cadmiumred}{#1}}
\newcommand{\bgreen}[1]{\textbf{\green{#1}}}

\newcommand{\cutblur}{CutBlur}
\newcommand{\Fref}[1]{Figure \ref{#1}}
\newcommand{\Sref}[1]{Section \ref{#1}}
\newcommand{\Tref}[1]{Table \ref{#1}}

\newcommand{\customfootnotetext}[2]{
    {\renewcommand{\thefootnote}{#1}\footnotetext[0]{#2}}
}

\newcommand{\minus}{\hspace{0.23ex}-\hspace{0.22ex}}

\usepackage[pagebackref=true,breaklinks=true,letterpaper=true,colorlinks,bookmarks=false]{hyperref}

\cvprfinalcopy 

\def\cvprPaperID{7962} 

\begin{document}

\title{Rethinking Data Augmentation for Image Super-resolution:\\A Comprehensive Analysis and a New Strategy}

\author{Jaejun Yoo\footnotemark[1]\\
EPFL.\\
{\tt\small jaejun.yoo88@gmail.com}
\and
Namhyuk Ahn\footnotemark[1]\\
Ajou University\\
{\tt\small aa0dfg@ajou.ac.kr}
\and
Kyung-Ah Sohn\textsuperscript{$\dagger$}\\
Ajou University\\
{\tt\small kasohn@ajou.ac.kr}
}

\maketitle

\begin{abstract}
   Data augmentation is an effective way to improve the performance of deep networks. Unfortunately, current methods are mostly developed for high-level vision tasks (\eg, classification) and few are studied for low-level vision tasks (\eg, image restoration). In this paper, we provide a comprehensive analysis of the existing augmentation methods applied to the super-resolution task. We find that the methods discarding or manipulating the pixels or features too much hamper the image restoration, where the spatial relationship is very important. Based on our analyses, we propose \textbf{\cutblur} that cuts a low-resolution patch and pastes it to the corresponding high-resolution image region and vice versa. The key intuition of \cutblur~is to enable a model to learn not only ``how" but also ``where" to super-resolve an image. By doing so, the model can understand ``how much", instead of blindly learning to apply super-resolution to every given pixel. Our method consistently and significantly improves the performance across various scenarios, especially when the model size is big and the data is collected under real-world environments. We also show that our method improves other low-level vision tasks, such as denoising and compression artifact removal.
\end{abstract}
\vspace{-0.5 cm}
\section{Introduction}
\customfootnotetext{*}{~indicates equal contribution. Most work was done in NAVER Corp.}
\customfootnotetext{$\dagger$}{~indicates corresponding author.}

Data augmentation (DA) is one of the most practical ways to enhance model performance without additional computation cost in the test phase.
While various DA methods~\cite{cutout, cutmix, mixup, imagenet-c} have been proposed in several high-level vision tasks, DA in low-level vision has been scarcely investigated. Instead, many image restoration studies, such as super-resolution (SR), have relied on the synthetic datasets~\cite{timofte2017ntire}, which we can easily increase the number of training samples by simulating the system degradation functions (\eg, using the bicubic kernel for SR).

Because of the gap between a simulated and a real data distribution, however, models that are trained on simulated datasets do not exhibit optimal performance in the real environments~\cite{realsr}. Several recent studies have proposed to mitigate the problem by collecting real-world datasets~\cite{sidd,realsr,zhang2018single}. 
However, in many cases, it is often very time-consuming and expensive to obtain a large number of such data. 
Although this is where DA can play an important role, only a handful of studies have been performed~\cite{overfit-sr,seven-radu}.

Radu \etal~\cite{seven-radu} was the first to study various techniques to improve the performance of example-based single-image super-resolution (SISR), one of which was data augmentation. Using rotation and flipping, they reported consistent improvements across models and datasets. Still, they only studied simple geometric manipulations with traditional SR models~\cite{sc, a+} and a very shallow learning-based model, SRCNN~\cite{srcnn}. 
To the best of our knowledge, Feng \etal~\cite{overfit-sr} is the only work that analyzed a recent DA method (Mixup~\cite{mixup}) in the example-based SISR problem. 
However, the authors provided only a limited observation using a single U-Net-like architecture and tested the method with a single dataset (RealSR~\cite{realsr}).

\begin{figure*}[t]
\centering
\subfigure[High resolution]{\includegraphics[width=0.16\linewidth]{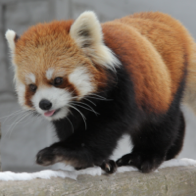}}
\subfigure[Low resolution]{\includegraphics[width=0.16\linewidth]{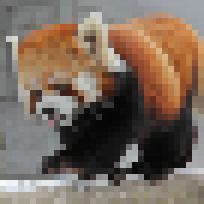}}
\subfigure[\cutblur]{\includegraphics[width=0.16\linewidth]{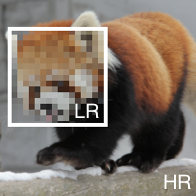}} \hspace{0.96em}
\subfigure[Schematic illustration of \cutblur~operation]{\includegraphics[width=0.465\linewidth]{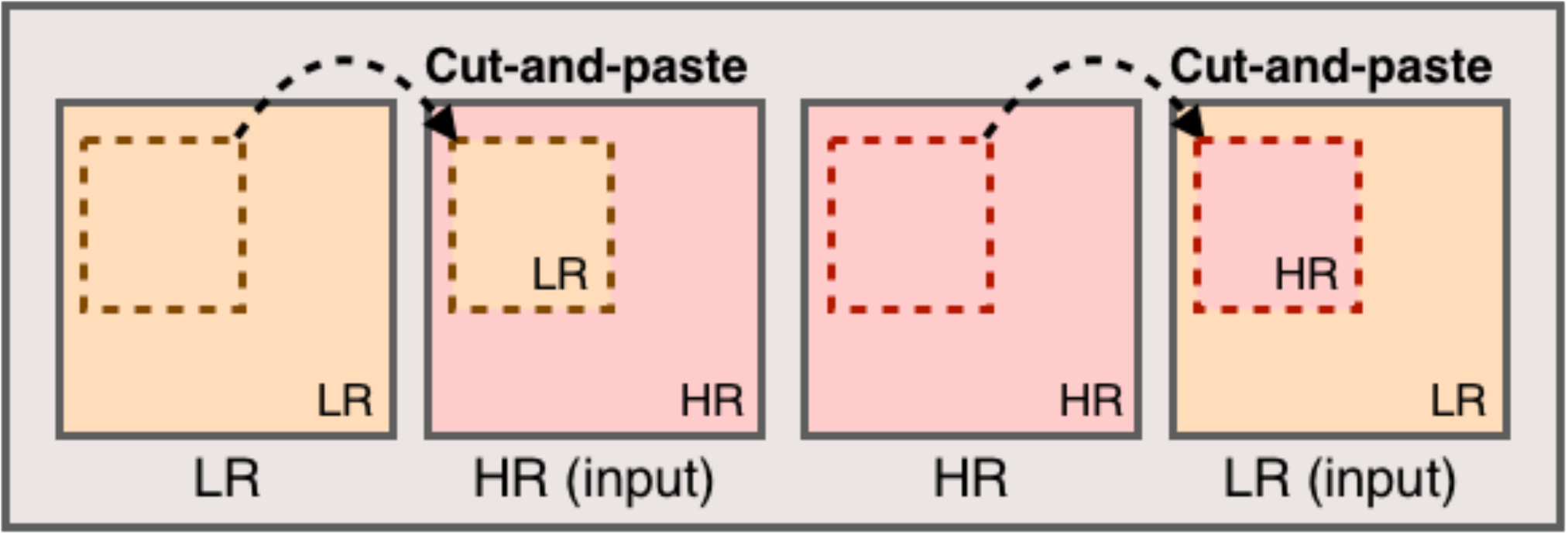}}\\\vspace{-0.5em}
\subfigure[Blend]{\includegraphics[width=0.16\linewidth]{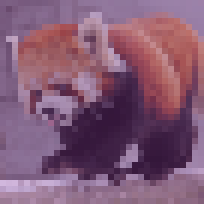}}
\subfigure[RGB permute]{\includegraphics[width=0.16\linewidth]{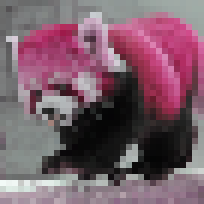}}
\subfigure[Cutout (25\%)~\cite{cutout}]{\includegraphics[width=0.16\linewidth]{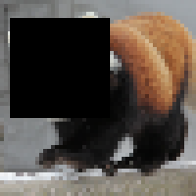}}
\subfigure[Mixup~\cite{mixup}]{\includegraphics[width=0.16\linewidth]{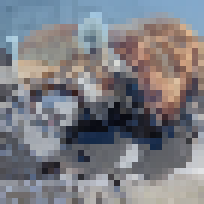}}
\subfigure[CutMix~\cite{cutmix}]{\includegraphics[width=0.16\linewidth]{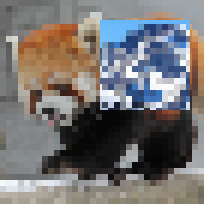}}
\subfigure[CutMixup]{\includegraphics[width=0.16\linewidth]{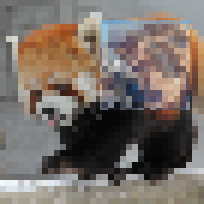}}\\
\caption{Data augmentation methods. \textbf{(Top)} An illustrative example of our proposed method, \cutblur. \cutblur~generates an augmented image by cut-and-pasting the low resolution (LR) input image onto the ground-truth high resolution (HR) image region and vice versa (\Sref{sec:cutblur}). 
\textbf{(Bottom)} Illustrative examples of the existing augmentation techniques and a new variation of CutMix and Mixup, CutMixup.}
\label{fig:overview}
\end{figure*}

To better understand DA methods in low-level vision, we provide a comprehensive analysis on the effect of various DA methods that are originally developed for high-level vision tasks (\Sref{sec:analysis}). We first categorize the existing augmentation techniques into two groups depending on where the method is applied; pixel-domain~\cite{cutout, cutmix, mixup} and feature-domain~\cite{dropblock, shakeshake, manifold-mixup, shakedrop}. When directly applied to SISR, we find that some methods harm the image restoration results and even hampers the training, especially when a method largely induces the loss or confusion of spatial information between nearby pixels (\eg, Cutout \cite{cutout} and feature-domain methods). Interestingly, basic manipulations like RGB permutation that do not cause a severe spatial distortion provide better improvements than the ones which induce unrealistic patterns or a sharp transition of the structure (\eg, Mixup \cite{mixup} and CutMix \cite{cutmix}). 


Based on our analyses, we propose \textbf{\cutblur}, a new augmentation method that is specifically designed for the low-level vision tasks. \cutblur~cut and paste a low resolution (LR) image patch into its corresponding ground-truth high resolution (HR) image patch 
(\Fref{fig:overview}). 
By having partially LR and partially HR pixel distributions with a random ratio in a single image, \cutblur~enjoys the regularization effect by encouraging a model to learn both \textit{``how"} and \textit{``where"} to super-resolve the image. 
One nice side effect of this is that the model also learns \textit{``how much"} it should apply super-resolution on every local part of a given image. While trying to find a mapping that can simultaneously maintain the input HR region and super-resolve the other LR region, the model adaptively learns to super-resolve an image. 
 
Thanks to this unique property, \cutblur~prevents over-sharpening of SR models, which can be commonly found in real-world applications (\Sref{subsec:wild}). 
In addition, we show that the performance can be further boosted by applying several curated DA methods together during the training phase, which we call \textbf{mixture of augmentations (MoA)} (\Sref{sec:cutblur}).
Our experiments demonstrate that the proposed strategy significantly and consistently improves the model performance over various models and datasets. 
Our contributions are summarized as follows:
\begin{enumerate}
    \item To the best of our knowledge, we are the first to provide comprehensive analysis of recent data augmentation methods when directly applied to the SISR task. 
    \item We propose a new DA method, \textbf{\cutblur}, which can reduce unrealistic distortions by regularizing a model to learn not only ``how" but also ``where" to apply the super-resolution to a given image. 
    \item Our mixed strategy shows consistent and significant improvements in the SR task, \textbf{achieving state-of-the-art (SOTA) performance in RealSR} \cite{realsr}. 
\end{enumerate}

\section{Data augmentation analysis}
\label{sec:analysis}
In this section, we analyze existing augmentation methods and compare their performances when applied to EDSR
\cite{edsr}, which is our baseline super-resolution model. We train EDSR from scratch with DIV2K~\cite{div2k} dataset or RealSR~\cite{realsr} dataset. We used the authors' official code. 

\subsection{Prior arts}
\noindent\textbf{DA in pixel space.} There have been many studies to augment images in high-level vision tasks~\cite{cutout, cutmix, mixup} (\Fref{fig:overview}). 
Mixup~\cite{mixup} blends two images to generate an unseen training sample. %
Cutout and its variants~\cite{cutout,random-erase} drop a randomly selected region of an image. Addressing that Cutout cannot fully exploit the training data, CutMix~\cite{cutmix} replaces the random region with another image. 
Recently, AutoAugment and its variant \cite{autoaug,fastautoaug} have been proposed to learn the best augmentation policy for a given task and dataset.

\smallskip

\noindent\textbf{DA in feature space.}
DA methods manipulating CNN features have been proposed~\cite{adl, shakeshake, dropblock, dropout,manifold-mixup, shakedrop} and can be categorized into three groups: 1) feature mixing, 2) shaking, and 3) dropping.  Like Mixup, Manifold Mixup~\cite{manifold-mixup} mixes both input image and the latent features. 
Shake-shake~\cite{shakeshake} and ShakeDrop~\cite{shakedrop} 
perform a stochastic affine transformation to the features. 
Finally, following the spirit of Dropout~\cite{dropout}, a lot of feature dropping strategies~\cite{adl, dropblock, spatialdropout} have been proposed to boost the generalization of a model.
\smallskip

\noindent\textbf{DA in super-resolution.} A simple geometric manipulation, such as rotation and flipping, has been widely used in SR models \cite{seven-radu}. Recently, Feng \etal~\cite{overfit-sr} showed that Mixup can alleviate the overfitting problem of SR models \cite{realsr}.

\subsection{Analysis of existing DA methods}
\label{sec:existing-da}
The core idea of many augmentation methods is to partially block or confuse the training signal so that the model acquires more generalization power. However, unlike the high-level tasks, such as classification, where a model should learn to abstract an image, the local and global relationships among pixels are especially more important in the low-level vision tasks, such as denoising and super-resolution. Considering this characteristic, it is unsurprising that DA methods, which lose the spatial information, limit the model's ability to restore an image. 
Indeed, we observe that the methods dropping the information~\cite{adl,dropblock, spatialdropout} are detrimental to the SR performance and especially harmful in the feature space, which has larger receptive fields. Every feature augmentation method significantly drops the performance. Here, we put off the results of every DA method that degrades the performance in the supplementary material.  

On the other hand, DA methods in pixel space bring some improvements when applied carefully (\Tref{table:da})\footnote{For every experiment, we only used geometric DA methods, flip and rotation, which is the default setting of EDSR. Here, to solely analyze the effect of the DA methods, we did not use the $\times2$ pre-trained model.}. For example, Cutout~\cite{cutout} with default setting (dropping 25\% of pixels in a rectangular shape) significantly degrades the original performance by 0.1 dB. However, we find that Cutout gives a positive effect (DIV2K: +$0.01$ dB and RealSR: +$0.06$ dB) when applied with 0.1\% ratio and erasing random pixels instead of a rectangular region. Note that this drops only 2$\sim$3 pixels when using a 48$\times$48 input patch.

CutMix~\cite{cutmix} shows a marginal improvement (\Tref{table:da}), and we hypothesize that this happens because CutMix generates a drastically sharp transition of image context making boundaries. Mixup improves the performance but it mingles the context of two different images, which can confuse the model.
To alleviate these issues, we create a variation of CutMix and Mixup, which we call CutMixup (below the dashed line, \Fref{fig:overview}). Interestingly, it gives a better improvement on our baseline. By getting the best of both methods, CutMixUp benefits from minimizing the boundary effect as well as the ratio of the mixed contexts. 

Based on these observations, we further test a set of basic operations such as RGB permutation and Blend (adding a constant value) that do not incur any structural change in an image. (For more details, please see our supplementary material.) These simple methods show promising results in the synthetic DIV2K dataset and a big improvement in the RealSR dataset, which is more difficult. These results empirically prove our hypothesis, which naturally leads us to a new augmentation method, \cutblur. When applied, \cutblur~not only improves the performance (\Tref{table:da}) but provides some good properties and synergy (\Sref{sec:cutblur-discussion}), which cannot be obtained by the other DA methods. 

\begin{table}[!tb]
\centering
\caption{PSNR (dB) comparison of different data augmentation methods in super-resolution. We report the baseline model (EDSR \cite{edsr}) performance that is trained on DIV2K ($\times 4$)~\cite{div2k} and RealSR ($\times 4$)~\cite{realsr}. The models are trained from scratch. $\delta$ denotes the performance gap between with and without augmentation.}
\vspace{+0.1 cm}
\centering
\begin{tabular}{c|cc}
\hline
\multirow{2}{*}{Method} & \multirow{2}{*}{DIV2K ($\delta$)} & \multirow{2}{*}{RealSR ($\delta$)} \\\\
\hline
EDSR                   & 29.21 (+0.00) & 28.89 (+0.00) \\
\hline
Cutout~\cite{cutout} (0.1\%)         & 29.22 \green{(+0.01)} & 28.95 \green{(+0.06)} \\
CutMix~\cite{cutmix}   & 29.22 \green{(+0.01)} & 28.89 (+0.00)  \\                            
Mixup~\cite{mixup}     & 29.26 \green{(+0.05)} & 28.98 \green{(+0.09)} \\               
\cdashline{1-3}
CutMixup               & 29.27 \green{(+0.06)} & 29.03 \green{(+0.14)} \\
RGB perm.              & 29.30 \green{(+0.09)} & 29.02 \green{(+0.13)} \\                           
Blend                  & 29.23 \green{(+0.02)} & 29.03 \green{(+0.14)} \\               
\hline\hline
\cutblur               & \textbf{29.26 \green{(+0.05)}} & \textbf{29.12 \green{(+0.23)}} \\ 
All DA's (random)           & \textbf{29.30 \green{(+0.09)}} & \textbf{29.16 \green{(+0.27)}} \\
\hline       \end{tabular}\label{table:da}
\end{table}

\begin{figure*}[t]
\centering
\includegraphics[width=1.0\linewidth]{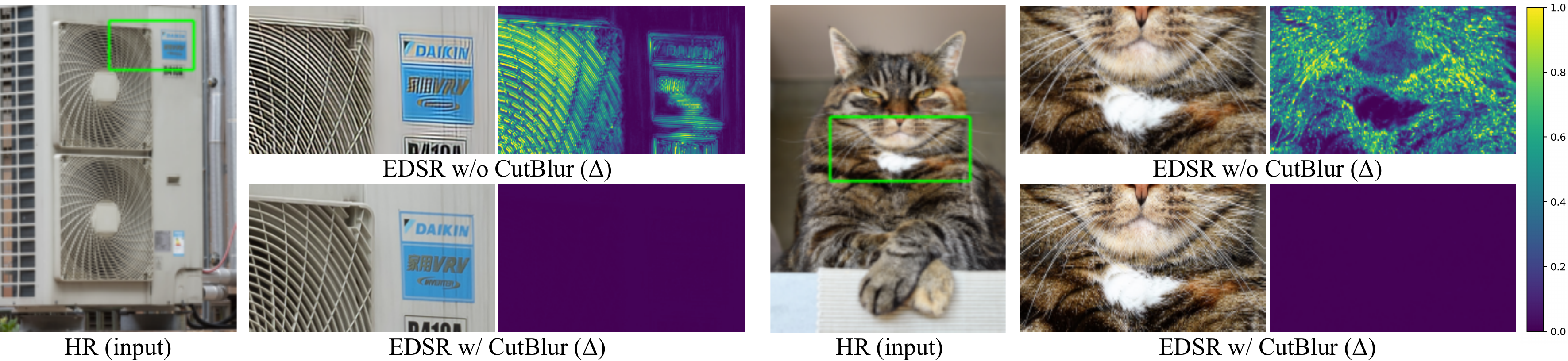}
\caption{Qualitative comparison of the baseline with and without \cutblur~when the network takes the HR image as an input during the inference time. $\Delta$ is the absolute residual intensity map between the network output and the ground-truth HR image. \cutblur~successfully preserves the entire structure while the baseline generates unrealistic artifacts (\textbf{left}) or incorrect outputs (\textbf{right}).} 
\vspace{-0.3 cm}
\label{fig:cutblur}
\end{figure*}

\section{\cutblur}
\label{sec:cutblur}
In this section, we describe the \cutblur, a new augmentation method that is designed for the super-resolution task.
\subsection{Algorithm}
Let $x_{LR}\in \mathbb{R}^{W\times H\times C}$ and $x_{HR}\in \mathbb{R}^{sW\times sH\times C}$ are LR and HR image patches and $s$ denotes a scale factor in the SR. As illustrated in \Fref{fig:overview}, because \cutblur~requires to match the resolution of $x_{LR}$ and $x_{HR}$, we first upsample $x_{LR}$ by $s$ times using a bicubic kernel, $x^s_{LR}$. The goal of \cutblur~is to generate a pair of new training samples $(\hat{x}_{HR\rightarrow LR}, \hat{x}_{LR\rightarrow HR})$ by cut-and-pasting the random region of $x_{HR}$ into the corresponding $x^s_{LR}$ and vice versa:
\begin{equation}
\begin{aligned}
\hat{x}_{HR\rightarrow LR} &= \mathbf{M}\odot x_{HR} + (\mathbf{1}-\mathbf{M})\odot x^s_{LR} \\
\hat{x}_{LR\rightarrow HR} &= \mathbf{M}\odot x^s_{LR} + (\mathbf{1}-\mathbf{M})\odot x_{HR}
\end{aligned}
\end{equation}
where $\mathbf{M}\in\{0,1\}^{sW\times sH}$ denotes a binary mask indicating where to replace, $\mathbf{1}$ is a binary mask filled with ones, and $\odot$ is element-wise multiplication. For sampling the mask and its coordinates, we follow the original CutMix \cite{cutmix}. 

\subsection{Discussion}
\label{sec:cutblur-discussion}
\noindent\textbf{Why \cutblur~works for SR?} 
In the previous analysis (\Sref{sec:existing-da}), we found that sharp transitions or mixed image contents within an image patch, or losing the relationships of pixels can degrade SR performance. 
Therefore, a good DA method for SR should not make unrealistic patterns or information loss while it has to serve as a good regularizer to SR models. 

\cutblur~satisfies these conditions because it performs cut-and-paste between the LR and HR image patches of the same content. By putting the LR (resp. HR) image region onto the corresponding HR (resp. LR) image region, it can minimize the boundary effect, which majorly comes from a mismatch between the image contents (\eg, Cutout and CutMix). Unlike Cutout, \cutblur~can utilize the entire image information while it enjoys the regularization effect due to the varied samples of random HR ratios and locations. 
\smallskip

\begin{figure}[t!]
\centering
\includegraphics[width=\linewidth]{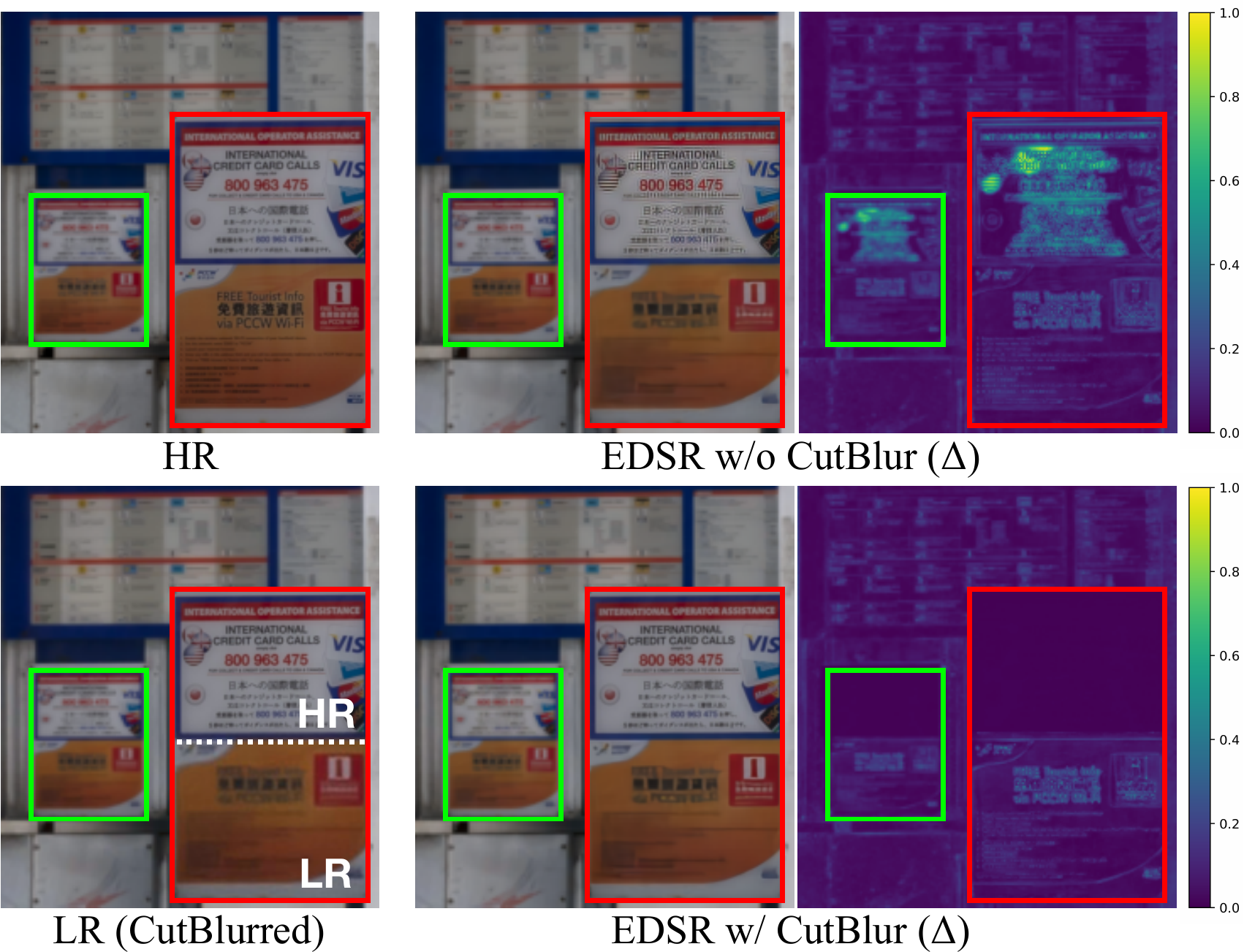}
\caption{Qualitative comparison of the baseline and \cutblur~model outputs when the input is augmented by \cutblur. $\Delta$ is the absolute residual intensity map between the network output and the ground-truth HR image. Unlike the baseline (\textbf{top right}), \cutblur~model not only resolves the HR region but reduces $\Delta$ of the other LR input area as well (\textbf{bottom right}).}
\vspace{-0.3 cm}
\label{fig:cutblur2}
\end{figure}

\noindent\textbf{What does the model learn with \cutblur?} 
Similar to the other DA methods that prevent classification models from over-confidently making a decision (\eg, label smoothing~\cite{labelsmoothing}), \cutblur~prevents the SR model from over-sharpening an image and helps it to super-resolve  only the necessary region. This can be demonstrated by performing the experiments with some artificial setups, where we provide the \cutblur-trained SR model with an HR image (\Fref{fig:cutblur}) or CutBlurred LR image (\Fref{fig:cutblur2}) as input.

When the SR model takes HR images at the test phase, it commonly outputs over-sharpened predictions, especially where the edges are (\Fref{fig:cutblur}). \cutblur~can resolve this issue by directly providing such examples to the model during the training phase. Not only does \cutblur~mitigate the over-sharpening problem, but it enhances the SR performance on the other LR regions, thanks to the regularization effect (\Fref{fig:cutblur2}). Note that the residual intensity has significantly decreased in the \cutblur~model. We hypothesize that this enhancement comes from constraining the SR model to discriminatively apply super-resolution to the image. Now the model has to simultaneously learn both ``how" and ``where" to super-resolve an image, and this leads the model to learn ``how much" it should apply super-resolution, which provides a beneficial regularization effect to the training. 

Of course it is unfair to compare the models that have been trained with and without such images. However, we argue that these scenarios are not just the artificial experimental setups but indeed exist in the real-world (\eg, out-of-focus images). We will discuss this more in detail with several real examples in \Sref{subsec:wild}.

\smallskip

\noindent\textbf{\cutblur~vs. Giving HR inputs during training.}
To make a model learn an identity mapping, instead of using \cutblur, one can easily think of providing HR images as an input of the network during the training phase. With the EDSR model, 
\cutblur~trained model (29.04 dB) showed better performance in PSNR than na\"ively providing the HR images (28.87 dB) to the network. (The detailed setups can be found in the supplementary material.) This is because \cutblur~is more general in that HR inputs are its special case ($\mathbf{M}=\mathbf{0}$ or $\mathbf{1}$). On the other hand, giving HR inputs can never simulate the mixed distribution of LR and HR pixels so that the network can only learn ``how", not ``where" to super-resolve an image. 
\smallskip

\noindent\textbf{Mixture of augmentation (MoA).} 
To push the limits of performance gains, we integrate various DA methods into a single framework. For each training iteration, the model first decides with probability $p$ whether to apply DA on inputs or not. If yes, it randomly selects a method among the DA pool. Based on our analysis, we use all the pixel-domain DA methods discussed in \Tref{table:da} while excluding all feature-domain DA methods. Here, we set $p=1.0$ as a default.  From now on, unless it is specified, we report all the experimental results using this MoA strategy.

\begin{figure}[t!]
\centering
\subfigure[SRCNN (0.07M)]{\includegraphics[height=8em,width=0.50\linewidth]{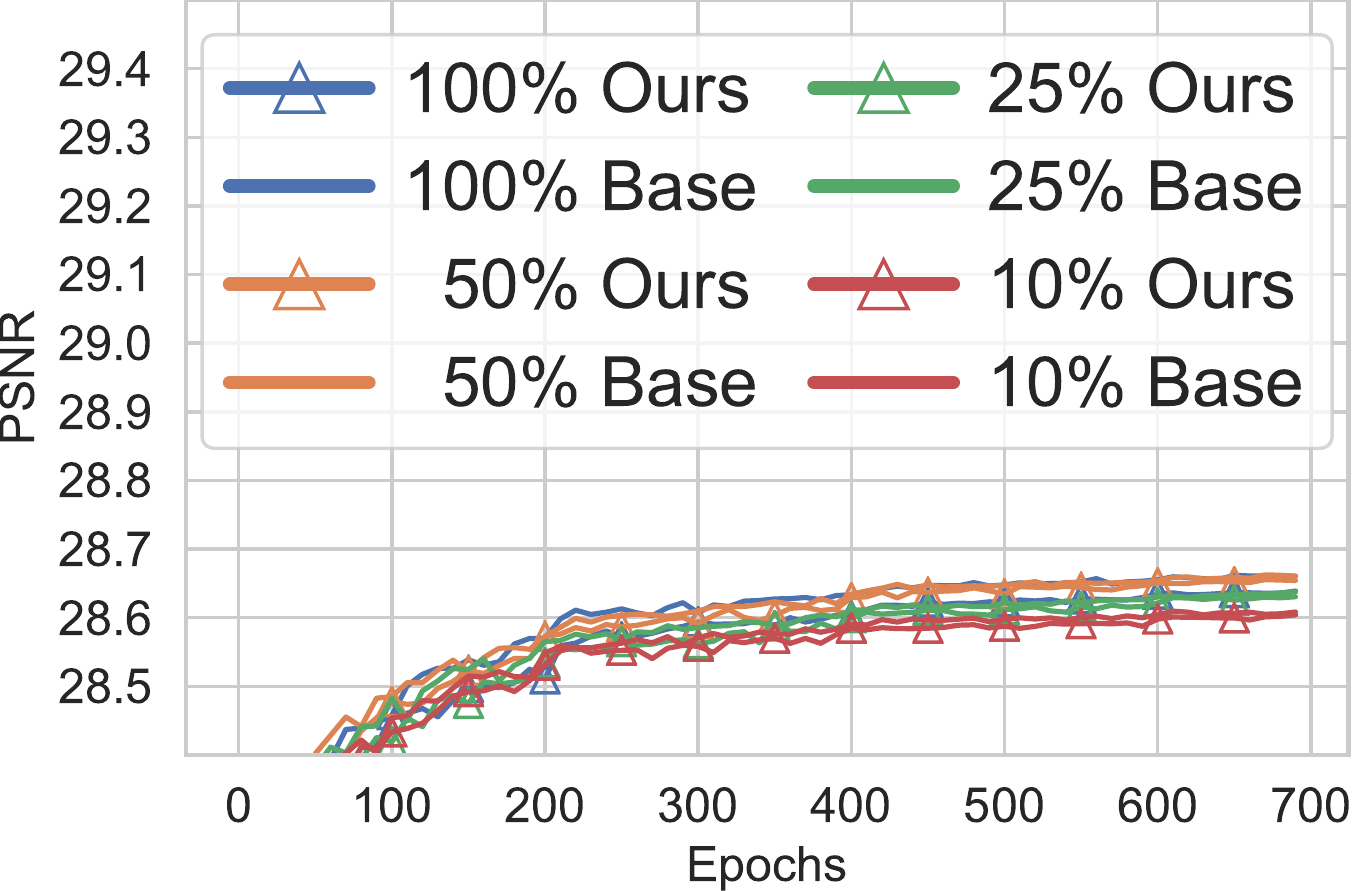}}
\subfigure[CARN (1.14M)]{\includegraphics[height=8em,width=0.46\linewidth]{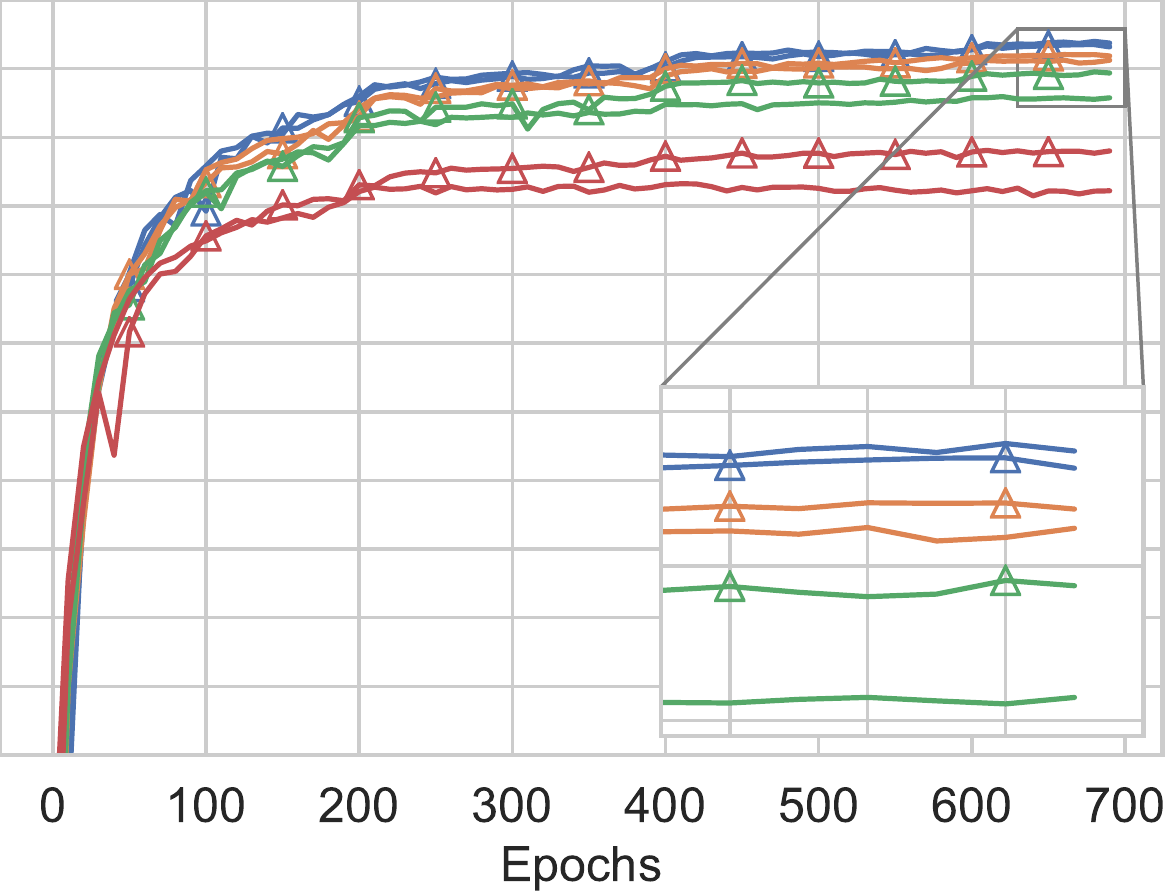}} \\ \vspace{-0.7em}
\subfigure[RCAN (15.6M)]{\includegraphics[height=8em,width=0.50\linewidth]{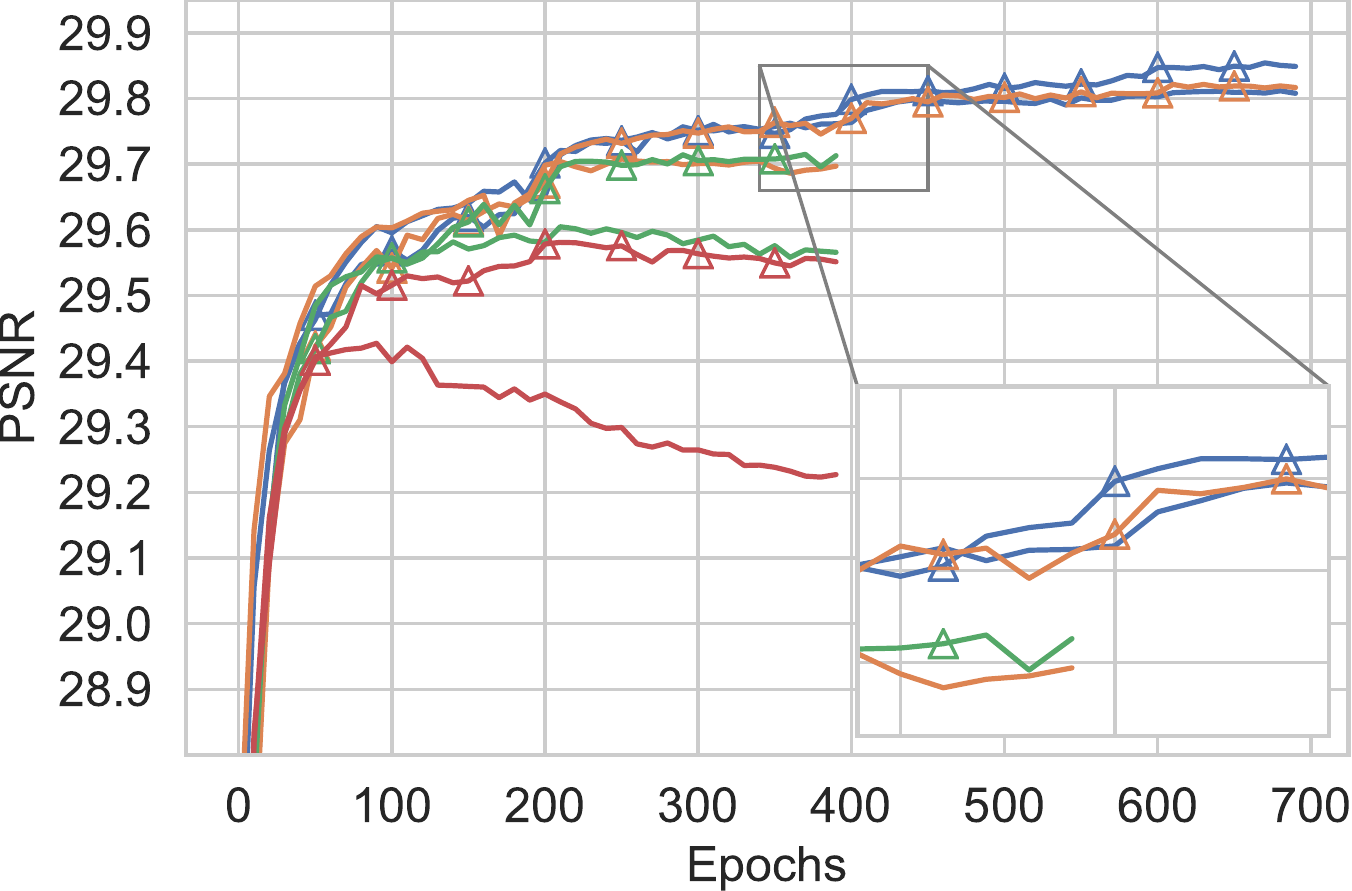}}
\subfigure[EDSR (43.2M)]{\includegraphics[height=8em,width=0.46\linewidth]{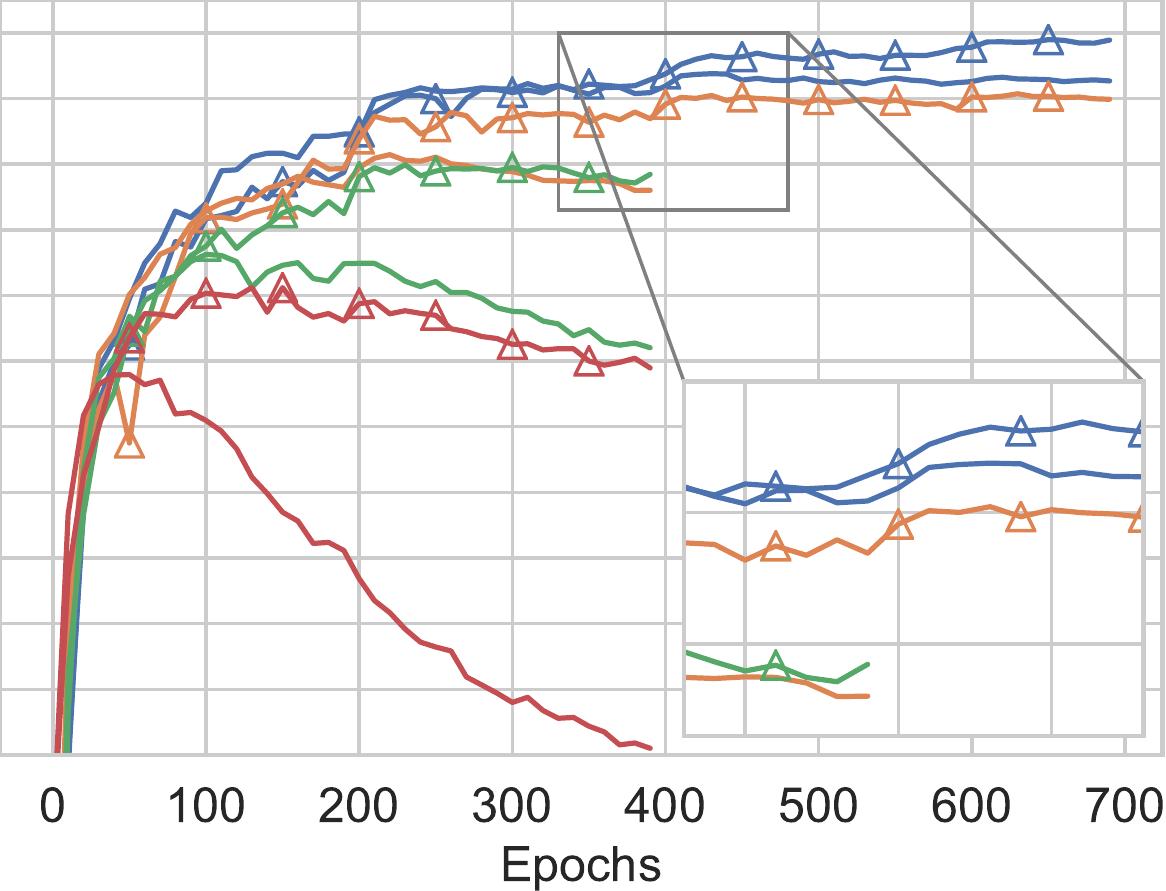}} \\
\caption{PSNR (dB) comparison on ten DIV2K ($\times 4$) validation images during training for different data size (\%). Ours are shown by triangular markers. Zoomed curves are displayed (inlets).}
\label{fig:curve}
\end{figure}

\begin{figure*}[t!]
\centering
\includegraphics[width=0.98\linewidth]{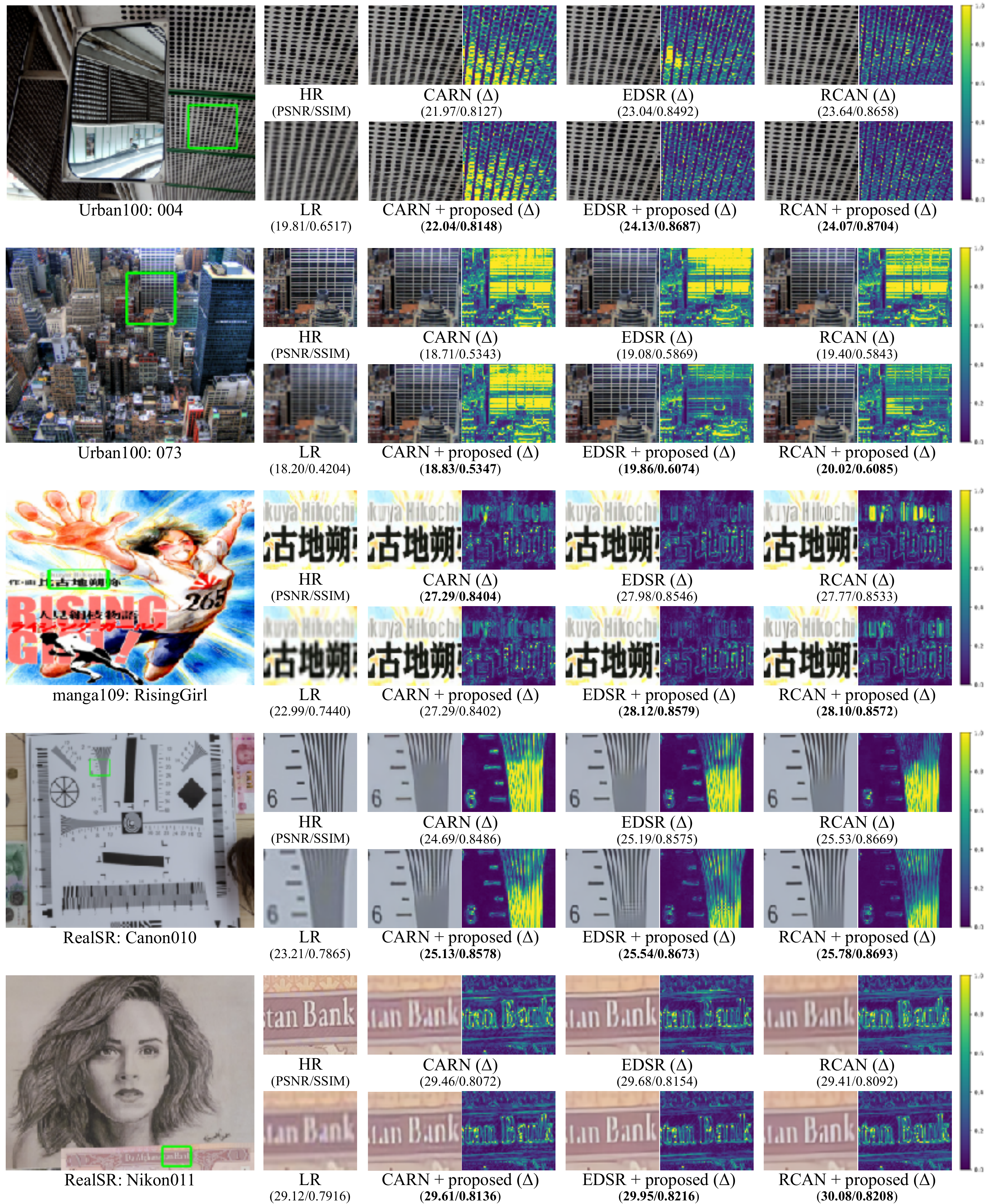}
\caption{Qualitative comparison of using our proposed method on different datasets and tasks.  $\Delta$ is the absolute residual intensity map between the network output and the ground-truth HR image. 
}
\label{fig:comp1}
\end{figure*}

\section{Experiments}
\label{sec:experiments}
In this section, we describe our experimental setups and compare the model performance with and without applying our method. We compare the super-resolution (SR) performance under various model sizes, dataset sizes (\Sref{subsec:size}), and benchmark datasets (\Sref{subsec:bm_comparison}). Finally, we apply our method to the other low-level vision tasks, such as Gaussian denoising and JPEG artifact removal, to show the potential extensibility of our method (\Sref{subsec:lowlevel}).\footnote{The overall experiments were conducted on NSML~\cite{sung2017nsml} platform.}\footnote{Our code is available at \href{https://github.com/clovaai/cutblur}{clovaai/cutblur}}
\smallskip

\noindent \textbf{Baselines.} We use four SR models: SRCNN~\cite{srcnn}, 
CARN~\cite{carn}, RCAN~\cite{rcan}, and EDSR \cite{edsr}. These models have different numbers of parameters from 0.07M to 43.2M (million).
For fair comparisons, every model is trained from scratch using the authors' official code unless mentioned otherwise.
\smallskip  

\noindent \textbf{Dataset and evaluation.} We use the DIV2K~\cite{div2k} dataset or a recently proposed real-world SR dataset, RealSR~\cite{realsr} for training. For evaluation, we use Set14~\cite{set14}, Urban100~\cite{urban100}, Manga109 \cite{manga109}, and test images of the RealSR dataset. Here, PSNR and SSIM are calculated on the Y channel only except the color image denoising task.

\subsection{Study on different models and datasets}
\label{subsec:size}
\noindent\textbf{Various model sizes.}
It is generally known that a large model benefits more from augmentation than a small model does. To see whether this is true in SR, we investigate how the model size affects the maximum performance gain using our strategy. 
Here, we set the probability of applying augmentations differently depending on the model size, $p=0.2$ for the small models (SRCNN and CARN) and $p=1.0$ for the large models (RCAN and EDSR). 
With the small models, our proposed method provides no benefit or marginally increases the performance (\Tref{table:comp_model}). This demonstrates the severe underfitting of the small models, where the effect of DA is minimal due to the lacking capacity. 
On the other hand, it consistently improves the performance of RCAN and EDSR, which have enough capacity to exploit the augmented information. 

\begin{table}[t!]
\centering
\small
\caption{PSNR (dB) comparison on DIV2K ($\times$4) validation set by varying the model and the size of dataset for training. Note that the number of RealSR dataset, which is more difficult to collect, is around 15\% of DIV2K dataset.}
\vspace{+0.1 cm}
\setlength{\tabcolsep}{4pt}
\centering
\begin{tabular}{l|c|ccc||cc}
\hline
\multirow{2}{*}{Model} & \multirow{2}{*}{Params.} & \multicolumn{5}{c}{Training Data Size}  \\
\cline{3-7}
  & & 100\% & 50\% & 25\% & \textbf{15\%} & \textbf{10\%} \\
\hline\hline
SRCNN     & \multirow{2}{*}{0.07M} & 27.95 & 27.95 & 27.95 & 27.93 & 27.91 \\
+ proposed    &                    & \red{\minus0.02} & \red{\minus0.01} & \red{\minus0.02} & \red{\minus0.02} & \red{\minus0.01} \\
\hline
CARN     & \multirow{2}{*}{1.14M} &  28.80 & 28.77 & 28.72 & 28.67 & 28.60 \\
+ proposed     &                  & +0.00 & \green{\textbf{+0.01}} & \green{\textbf{+0.02}} & \green{\textbf{+0.03}} & \green{\textbf{+0.04}} \\
\hline
RCAN     & \multirow{2}{*}{15.6M} & 29.22 & 29.06 & 29.01 & 28.90 & 28.82 \\
+ proposed    &                   & \green{\textbf{+0.08}} & \green{\textbf{+0.16}} & \green{\textbf{+0.11}} & \green{\textbf{+0.13}} & \green{\textbf{+0.14}}\\
\hline
EDSR     & \multirow{2}{*}{43.2M} & 29.21 & 29.10 & 28.97 & 28.87 & 28.77 \\
+ proposed     &                  & \green{\textbf{+0.08}} & \green{\textbf{+0.08}} & \green{\textbf{+0.10}} & \green{\textbf{+0.10}} & \green{\textbf{+0.11}} \\
\hline
\end{tabular}
\label{table:comp_model}
\end{table}

\begin{table*}[t]
\caption{Quantitative comparison (PSNR / SSIM) on SR (scale $\times$4) task in both synthetic and realistic settings. $\delta$ denotes the performance gap between with and without augmentation. For synthetic case, we perform the $\times$2 scale pre-training.}
\centering
\small
\begin{tabular}{l|c|ccc||c}
\hline
\multirow{2}{*}{Model} & \multirow{2}{*}{\# Params.} & \multicolumn{3}{c||}{Synthetic (DIV2K dataset)} & \multicolumn{1}{c}{Realistic (RealSR dataset)} \\
\cline{3-6}
&  & Set14 ($\delta$) & Urban100 ($\delta$) & Manga109 ($\delta$) & RealSR ($\delta$)\\
\hline\hline
CARN           & \multirow{2}{*}{1.14M} & 28.48 (+0.00) / 0.7787 & 25.85 (+0.00) / 0.7779 & 30.17 (+0.00) / 0.9034 & 28.78 (+0.00) / 0.8134  \\
+ proposed     &                        & 28.48 (+0.00) / 0.7788 & 25.85 (+0.00) / 0.7780 & 30.16 \red{(\minus0.01)} / 0.9032 & 29.00 \green{(+0.22)} / 0.8204 \\
\hline
RCAN           & \multirow{2}{*}{15.6M} & 28.86 (+0.00) / 0.7879 & 26.76 (+0.00) / 0.8062 & 31.24 (+0.00) / 0.9169 & 29.22 (+0.00) / 0.8254  \\
+ proposed     &                        & \textbf{28.92 \green{(+0.06)}} / \textbf{0.7895} & \textbf{26.93 \green{(+0.17)}} / \textbf{0.8106} & \textbf{31.46 \green{(+0.22)}} / \textbf{0.9190} & \textbf{29.49 \green{(+0.27)}} / \textbf{0.8307} \\
\hline
EDSR           & \multirow{2}{*}{43.2M} & 28.81 (+0.00) / 0.7871 & 26.66 (+0.00) / 0.8038 & 31.06 (+0.00) / 0.9151 & 28.89 (+0.00) / 0.8204 \\
+ proposed     &                        & 28.88 \green{(+0.07)} / 0.7886 & 26.80 \green{(+0.14)} / 0.8072 & 31.25 \green{(+0.19)} / 0.9163 & 29.16 \green{(+0.27)} / 0.8258 \\
\hline
\end{tabular}
\label{table:sr}
\end{table*}

\begin{figure*}[t!]
\centering
\includegraphics[height=50ex,width=1.0\linewidth]{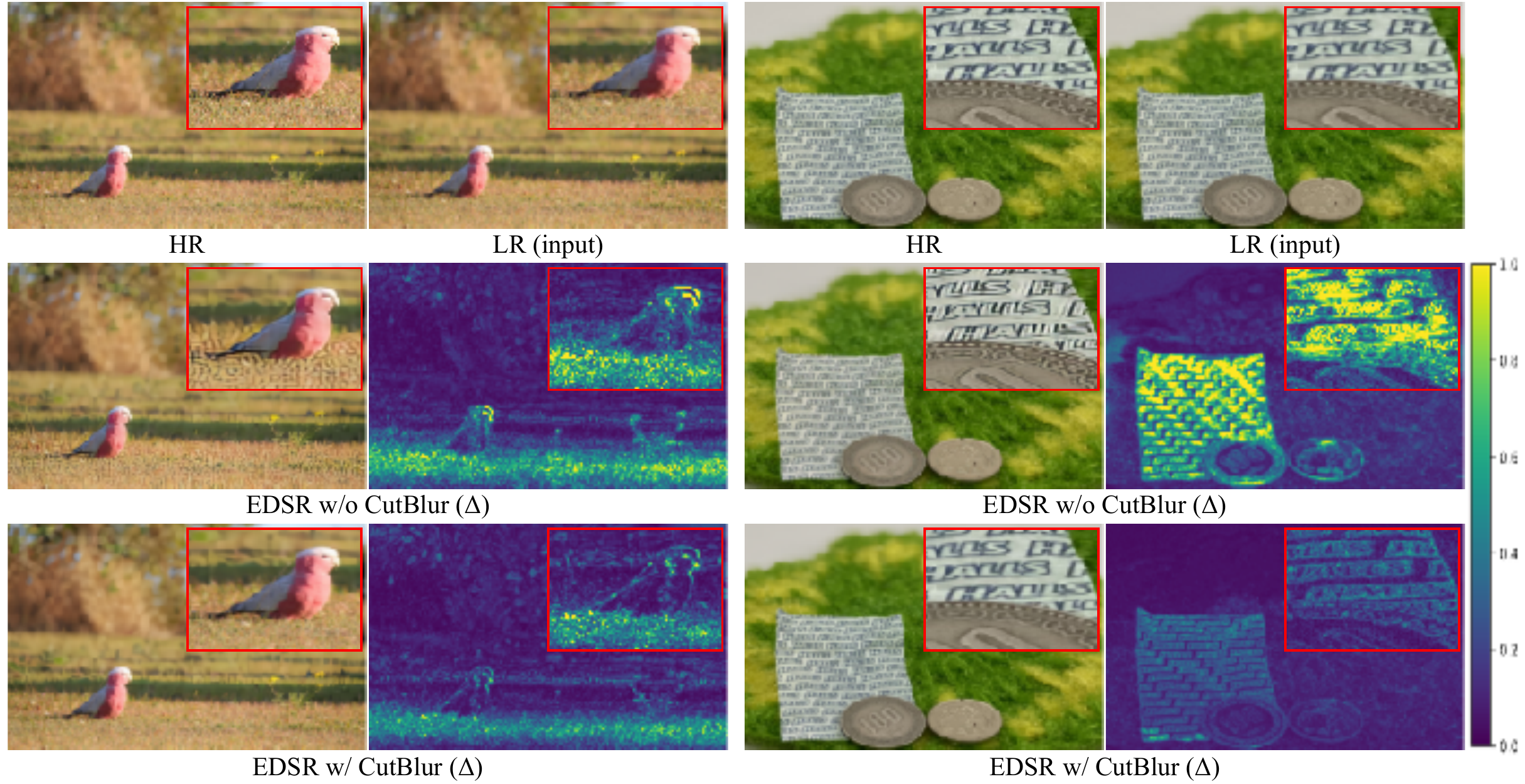}
\caption{Qualitative comparison of the baseline and \cutblur~model outputs. The inputs are the real-world out-of-focus photography ($\times2$ bicubic downsampled), which are taken from a web (\textbf{left}) and captured by iPhone 11 Pro (\textbf{right}).  $\Delta$ is the absolute residual intensity map between the network output and the ground-truth HR image.
The baseline model over-sharpens the focused region resulting in unpleasant distortions while the \cutblur~model effectively super-resolves the image without such a problem.}
\vspace{-0.3 cm}
\label{fig:outfocus}
\end{figure*}

\smallskip

\noindent\textbf{Various dataset sizes.}
We further investigate the model performance while decreasing the data size for training (\Tref{table:comp_model}). Here, we use 100\%, 50\%, 25\%, 15\% and 10\% of the DIV2K dataset. 
SRCNN and CARN show none or marginal improvements with our method. This can be also seen by the validation curves while training (\Fref{fig:curve}a and \ref{fig:curve}b). On the other hand, our method brings a huge benefit to the RCAN and EDSR in all the settings. The performance gap between the baseline and our method becomes profound as the dataset size diminishes. RCAN trained on half of the dataset shows the same performance as the 100\% baseline when applied with our method ($29.06 + 0.16 = 29.22$ dB). Our method gives an improvement of up to 0.16 dB when the dataset size is less than 50\%. This tendency is observed in EDSR as well. This is important because 15\% of the DIV2K dataset is similar to the size of the RealSR dataset, which is more expensive taken under real environments. 

Our method also significantly improves the overfitting problem (\Fref{fig:curve}c and \ref{fig:curve}d). 
For example, if we use 25\% of the training data, the large models easily overfit and this can be dramatically reduced by using our method (denoted by curves with the triangular marker of the same color).

\subsection{Comparison on diverse benchmark dataset}
\label{subsec:bm_comparison}
We test our method on various benchmark datasets.
For the synthetic dataset, we train the models using the DIV2K dataset and test them on Set14, Urban100, and Manga109. 
Here, we first pre-train the network with $\times$2 scale dataset, then fine-tune on $\times$4 scale images. For the realistic case, we train the models using the training set of the RealSR dataset ($\times$4 scale) and test them on its unseen test images.

Our proposed method consistently gives a huge performance gain, especially when the models have large capacities (\Tref{table:sr}). 
In the RealSR dataset, which is a more realistic case, the performance gain of our method becomes larger, \textbf{increasing at least 0.22 dB for all models in PSNR}. We achieve the \textbf{SOTA performance (RCAN~\cite{rcan})} compared to the previous SOTA model (LP-KPN~\cite{realsr}: 28.92 dB / 0.8340). Note that our model \textbf{increase the PSNR by 0.57 dB} with a comparable SSIM score. Surprisingly, \textbf{the lightest model (CARN~\cite{carn}: 1.14M) can already beat the LP-KPN (5.13M)} in PSNR with only $22\%$ of the parameters.

\Fref{fig:comp1} shows the qualitative comparison between the models with and without applying our DA method. In the Urban100 examples (1st and 2nd rows in \Fref{fig:comp1}), RCAN and EDSR benefit from the increased performance and successfully resolve the aliasing patterns. This can be seen more clearly in the residual between the model-prediction and the ground-truth HR image. Such a tendency is consistently observed across different benchmark images. In RealSR dataset images, even the performance of the small model is boosted, especially when there are fine structures (4th row in \Fref{fig:comp1}).


\subsection{\cutblur~in the wild}
\label{subsec:wild}
With the recent developments of devices like iPhone 11 Pro, they offer a variety of features, such as portrait images. Due to the different resolutions of the focused foreground and the out-focused background of the image, the baseline SR model shows degraded performance, while the  \cutblur~model does not (\Fref{fig:outfocus}). These are the very real-world examples, which are simulated by \cutblur. 
The baseline model adds unrealistic textures in the grass (left, \Fref{fig:outfocus}) and generates ghost artifacts around the characters and coin patterns (right, \Fref{fig:outfocus}). In contrast, the \cutblur~model does not add any unrealistic distortion while it adequately super-resolves the foreground and background of the image. 

\begin{table}[t!]
\caption{Performance comparison on the color Gaussian denoising task evaluated on the Kodak24 dataset. We train the model with both mild ($\sigma=30$) and severe noises ($\sigma=70$) and test on the mild setting. LPIPS \cite{lpips} (lower is better) indicates the perceptual distance between the network output and the ground-truth.}
\centering
\begin{tabular}{l|c|ccc}
\hline
\multirow{2}{*}{Model} & \multirow{2}{*}{Train $\sigma$} & \multicolumn{3}{c}{Test ($\sigma=30$)} \\
\cline{3-5}
& & PSNR $\uparrow$ & SSIM $\uparrow$ & LPIPS $\downarrow$\\
\hline\hline
EDSR     & \multirow{2}{*}{30} & 31.92 & \hspace{1ex}0.8716 & \hspace{1ex}0.136 \\
+ proposed & & \textbf{\green{+0.02}} & \textbf{\green{+0.0006}} & \textbf{\green{\minus0.004}} \\
\hline
EDSR     & \multirow{2}{*}{70} & 27.38 & \hspace{1ex}0.7295 & \hspace{1ex}0.375 \\
+ proposed & & \red{\minus2.51} & \textbf{\green{+0.0696}} & \textbf{\green{\minus0.193}} \\
\hline
\end{tabular}
\label{table:dn}
\end{table}

\begin{table}[t!]
\caption{Performance comparison on the color JPEG artifact removal task evaluated on the LIVE1~\cite{sheikh2005live} dataset. We train the model with both mild ($q=30$) and severe compression ($q=10$) and test on the mild setting.}
\centering
\begin{tabular}{l|c|ccc}
\hline
\multirow{2}{*}{Model} & \multirow{2}{*}{Train $q$} & \multicolumn{3}{c}{Test ($q=30$)} \\
\cline{3-5}
& & PSNR $\uparrow$ & SSIM $\uparrow$ & LPIPS $\downarrow$\\
\hline\hline
EDSR     & \multirow{2}{*}{30} & 33.95 & \hspace{1ex}0.9227 & \hspace{1ex}0.118 \\
+ proposed & & \red{\minus0.01} & \red{\minus0.0002} & \red{+0.001} \\
\hline
EDSR     & \multirow{2}{*}{10} & 32.45 & \hspace{1ex}0.8992 & \hspace{1ex}0.154 \\
+ proposed & & \textbf{\green{+0.97}} & \textbf{\green{+0.0187}} & \textbf{\green{\minus0.023}} \\
\hline
\end{tabular}
\label{table:jpeg}
\end{table}

\subsection{Other low-level vision tasks}
\label{subsec:lowlevel}
Interestingly, we find that our method also gives similar benefits when applied to the other low-level vision tasks. 
We demonstrate the potential advantages of our method by applying it to the Gaussian denoising and JPEG artifact removal tasks. 
For each task, we use EDSR as our baseline and trained the model from scratch with the synthetic DIV2K dataset with the corresponding degradation functions. We evaluate the model performance on the Kodak24 and LIVE1~\cite{sheikh2005live} datasets using PSNR (dB), SSIM, and LPIPS~\cite{lpips}.
Please see the appendix for more details.

\bigskip

\noindent\textbf{Gaussian denoising (color).}
We generate a synthetic dataset using Gaussian noise of different signal-to-noise ratios (SNR); $\sigma=$ 30 and 70 (higher $\sigma$ means stronger noise). Similar to the over-sharpening issue in SR, we simulate the over-smoothing problem (bottom row, \Tref{table:dn}). 
The proposed model has lower PSNR (dB) than the baseline but it shows higher SSIM and lower LPIPS~\cite{lpips}, which is known to measure the perceptual distance between two images (lower LPIPS means smaller perceptual difference). 

\begin{figure}[t!]
\centering
\includegraphics[height=40ex,width=\linewidth]{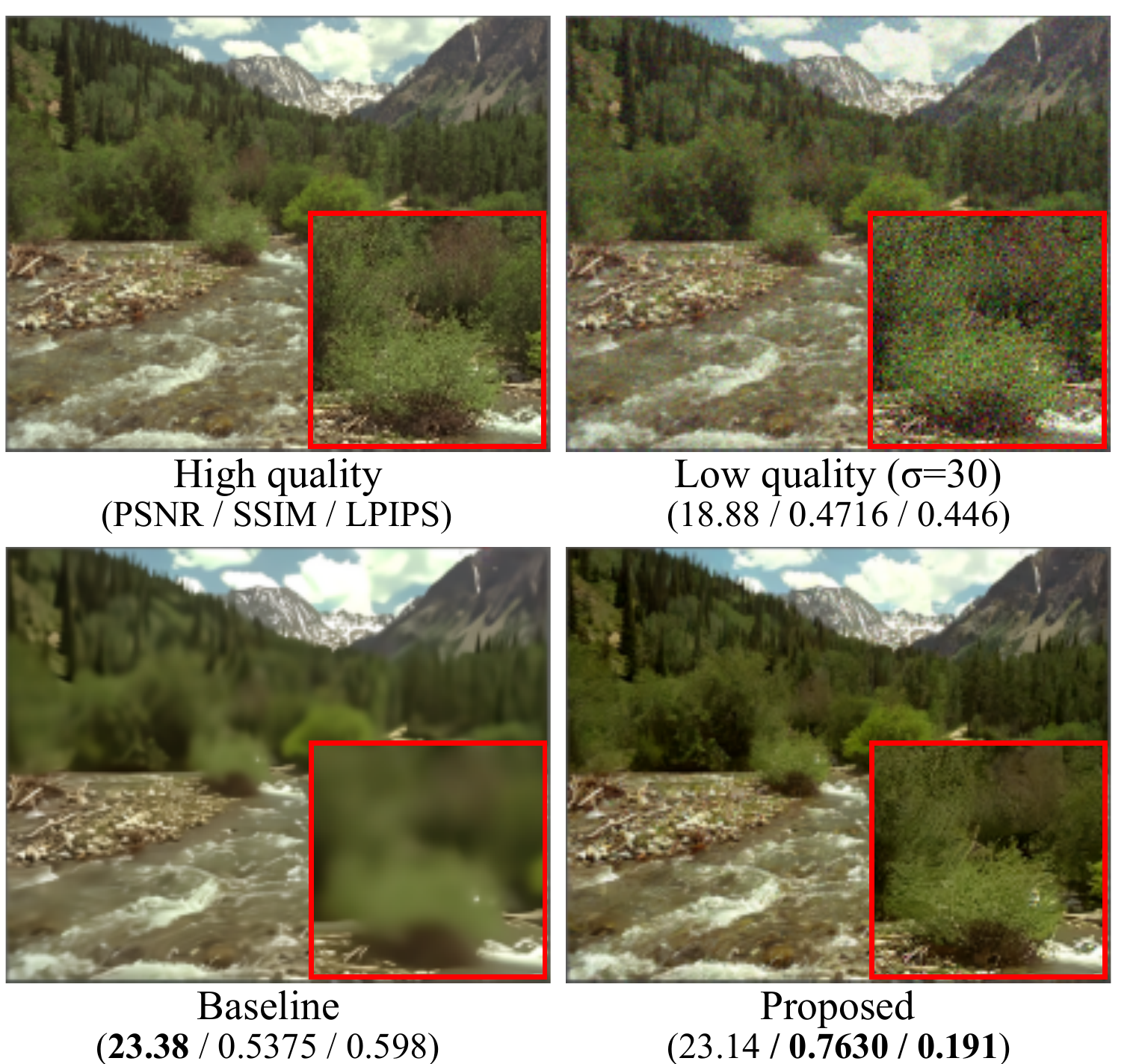}
\caption{Comparison of the generalization ability in the denoising task. Both the baseline and the proposed method are trained using $\sigma=70$ (severe) and tested with $\sigma=30$ (mild). Our proposed method effectively recovers the details while the baseline over-smooths the input resulting in a blurry image.}
\vspace{-0.3 cm}
\label{fig:dn}
\end{figure}

In fact, the higher PSNR of the baseline model is due to the over-smoothing
(\Fref{fig:dn}). Because the baseline model has learned to remove the stronger noise, it provides the over-smoothed output losing the fine details of the image. Due to this over-smoothing, its SSIM score is significantly lower and LPIPS is significantly higher. In contrast, the proposed model trained with our strategy successfully denoises the image while preserving the fine structures, which demonstrates the good regularization effect of our method.

\bigskip

\noindent\textbf{JPEG artifact removal (color).}
We generate a synthetic dataset using different compression qualities, $q=$ 10 and 30 (lower $q$ means stronger artifact) on the color image.
Similar to the denoising task, we simulate the over-removal issue. Compared to the baseline model, our proposed method shows significantly better performance in all metrics we used (bottom row, \Tref{table:jpeg}). The model generalizes better and gives 0.97 dB performance gain in PSNR.

\section{Conclusion}
We have introduced \cutblur~and Mixture of Augmentations (MoA), a new DA method and a strategy for training a stronger SR model. By learning how and where to super-resolve an image, \cutblur~encourages the model to understand how much it should apply the super-resolution to an image area. We have also analyzed which DA methods hurt SR performance and how to modify those to prevent such degradation. We showed that our proposed MoA strategy consistently and significantly improves the performance across various scenarios, especially when the model size is big and the dataset is collected from real-world environments. Last but not least, our method showed promising results in denoising and JPEG artifact removals, implying its potential extensibility to other low-level vision tasks. 

\appendix

\begin{table*}[!t]
\caption{A description of data augmentations that are used in our final proposed method.}
\centering
\bgroup
\def\arraystretch{1.2}
\begin{tabular}{l|lc}
Name & Description & Default $\alpha$ \\
\hline\hline
Cutout~\cite{cutout} & \multicolumn{1}{p{65ex}}{Erase (zero-out) randomly sampled pixels with probability $\alpha$. Cutout-ed pixels are discarded when calculating loss by masking removed pixels.} & 0.001 \\
CutMix~\cite{cutmix} & \multicolumn{1}{p{65ex}}{Replace randomly selected square-shape region to sub-patch from other image. The coordinates are calculated as: $r_x = \text{Unif}(0, W), \; r_w = \lambda W \; \text{, where} \; \lambda \sim \text{N}(\alpha, 0.01)$ (same for $r_y$ and $r_h$).} & 0.7 \\
Mixup~\cite{mixup} & \multicolumn{1}{p{65ex}}{Blend randomly selected two images. We use default setting of Feng \etal~\cite{overfit-sr} which is: $I' = \lambda I_i + (1-\lambda) I_j \; \text{, where} \; \lambda \sim \text{Beta}(\alpha, \alpha)$.} & 1.2 \\
CutMixup & \multicolumn{1}{p{65ex}}{CutMix with the Mixup-ed image. CutMix and Mixup procedure use hyper-parameter $\alpha_1$ and $\alpha_2$ respectively.} & \multicolumn{1}{p{8ex}}{0.7 / 1.2 ($\alpha_1$ / $\alpha_2$)} \\
RGB permutation & Randomly permute RGB channels. & - \\
Blend & \multicolumn{1}{p{65ex}}{Blend image with vector $\textbf{v} = (v_1, v_2, v_3) \; \text{, where} \; v_i \sim \text{Unif}(\alpha, 1)$.} & 0.6 \\
\hline\hline
CutBlur & \multicolumn{1}{p{65ex}}{Perform CutMix with same image but different resolution, producing $\hat{x}_{HR\rightarrow LR}$ and $\hat{x}_{LR\rightarrow HR}$. Randomly choose $\hat{x}$ from the [$\hat{x}_{HR\rightarrow LR}$, $\hat{x}_{LR\rightarrow HR}$], then provided selected one as input of the network.} & 0.7 \\
\makecell[l]{\\MoA \\ (Mixture of Augmentations)} & \multicolumn{1}{p{65ex}}{Use all data augmentation method described above. Randomly select single augmentation from the augmentation pool then apply it.} & - \\
\hline
\end{tabular}
\egroup
\label{table:da_setup}
\end{table*}

\medskip
\noindent{\textbf{Acknowledgement.} We would like to thank Clova AI Research team, especially Yunjey Choi, Seong Joon Oh, Youngjung Uh, Sangdoo Yun, Dongyoon Han, Youngjoon Yoo, and Jung-Woo Ha for their valuable comments and feedback. This work was supported by NAVER Corp and also by the National Research Foundation of Korea grant funded by the Korea government (MSIT) (no.NRF-2019R1A2C1006608)}

\newpage

\appendix

\section{Implementation Details}
\noindent\textbf{Network modification.}
To apply \cutblur, the resolution of the input and output has to match. To satisfy such requirement, we first upsample the input $x_{LR} \in \mathbb{R}^{W\times H\times C}$ to $x_{LR}^s \in \mathbb{R}^{sW\times sH\times C}$ using \textit{bicubic} kernel then feed it to the network. In order to achieve efficient inference, we attach \textit{desubpixel} layer~\cite{desubpixel} at the beginning of the network. By adapting this layer, input is reshaped as $x_{LR}^s \in \mathbb{R}^{W\times H\times s^2C}$ so that the entire forward pass is performed on the low resolution space. Note that such modifications are only for the synthetic SR task because the other low-level tasks (\textit{e.g.} denoising) have an identical input and output size.

Table~\ref{table:mod} shows the performance of original and modified networks. For both RCAN and EDSR, modified networks reach the performance of the original one with negligible increases in the number of the parameters and the inference time. Note that we measure the inference time on the NVIDIA V100 GPU using a resolution of 480$\times$320 for the LR input so that the network generates a 2K SR image.

\noindent\textbf{Augmentation setup.}
Detailed description and setting of every augmentation that we used are described in Table~\ref{table:da_setup}. Here, \textit{CutMixup}, \textit{CutBlur}, and \textit{MoA} are the strategies that we have newly proposed in the paper. The hyper-parameters are described following the original papers' notations.

Unless mentioned, at each iteration, we always apply MoA ($p=1.0$) and evenly choose one method from the augmentation pool.
However, we set $p=0.2$ for training SRCNN and CARN on the synthetic SR dataset and $p=0.6$ for all the other models for denoising and compression artifact removal tasks, \ie, MoA is applied less. For the realistic SR task (RealSR dataset), we adjust the ratio of MoA to have CutBlur 40\% chance more than the other DA's, each of which has 10\% chance ($40\%+10\%+10\%+10\%+10\%+10\%+10\%=100\%$).

\begin{table}[t]
\caption{Performance (PSNR) and the model size (\# parameters and inference time) comparison between the original (\textit{ori.}) and modified (\textit{mod.}) networks on $\times$4 scale SR dataset. We borrow the reported scores from the performance of the original networks.}
\vspace{+0.1 cm}
\small
\setlength{\tabcolsep}{5pt}
\centering
\begin{tabular}{l|cc|ccc}
\hline
Model & \# Params. & Time & Set14 & Urban & Manga \\\hline\hline
RCAN (ori.) & 15.6M & 0.612s & 28.87 & 26.82 & 31.22 \\
RCAN (mod.) & 15.6M & 0.614s & 28.86 & 26.76 & 31.24 \\\hline
EDSR (ori.) & 43.1M & 0.334s & 29.80 & 26.64 & 31.02 \\
EDSR (mod.) & 43.2M & 0.335s & 28.81 & 26.66 & 31.06 \\
\hline
\end{tabular}
\label{table:mod}
\end{table}

\noindent\textbf{Evaluation protocol.}
To evaluate the performance of the model, we use three metrics: peak-signal-to-noise ratio (PSNR), structural similarity index (SSIM)~\cite{wang2004image}, and learned perceptual image patch similarity (LPIPS)~\cite{lpips}. PSNR is defined using the maximum pixel value and mean-squared error between the two images in the log-space. SSIM~\cite{wang2004image} measures the structural similarity between two images based on the luminance, contrast and structure.
Note that we use Y channel only when calculating PSNR and SSIM unless otherwise specified.

Although high PSNR and high SSIM of an image are generally interpreted as a good image restoration quality, it is well known that these metrics cannot represent human visual perception very well \cite{lpips}. LPIPS~\cite{lpips} has been recently proposed to address this mismatch. It measures the diversity of the generated images using the L1 distance between features extracted from the pre-trained AlexNet \cite{alexnet}, which gives better perceptual distance between two images than the traditional metrics. For more details, please refer to the original paper \cite{lpips}.

\section{Detailed Analysis}
In this section, we describe each experiment that has been introduced in the Analysis Section. We also provide the results of feature augmentation methods and the original cutout that we excluded in the main text. 
\smallskip

\begin{figure}[t]
\centering
\subfigure[RCAN]{\includegraphics[width=0.48\linewidth]{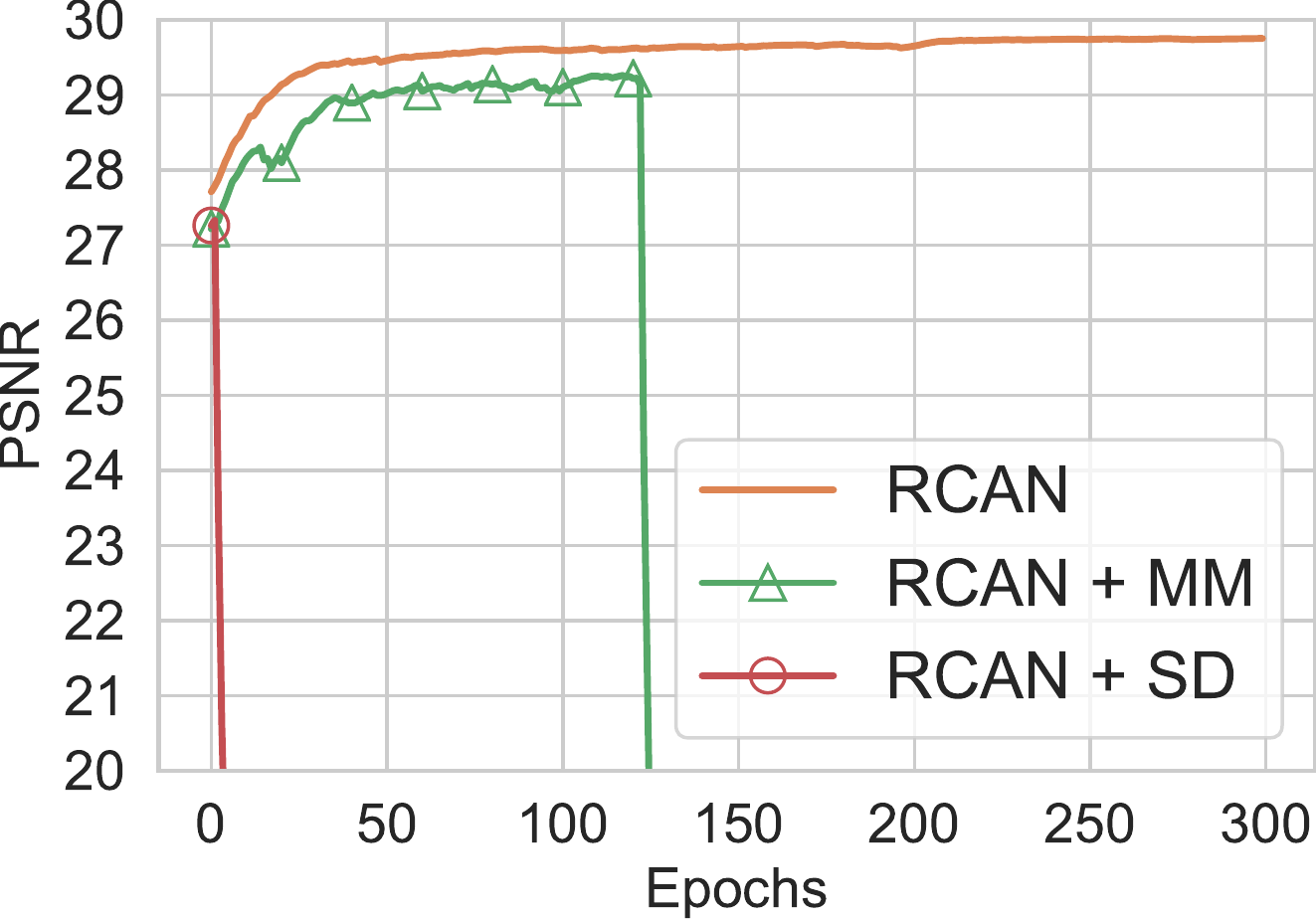}}
\subfigure[EDSR]{\includegraphics[width=0.48\linewidth]{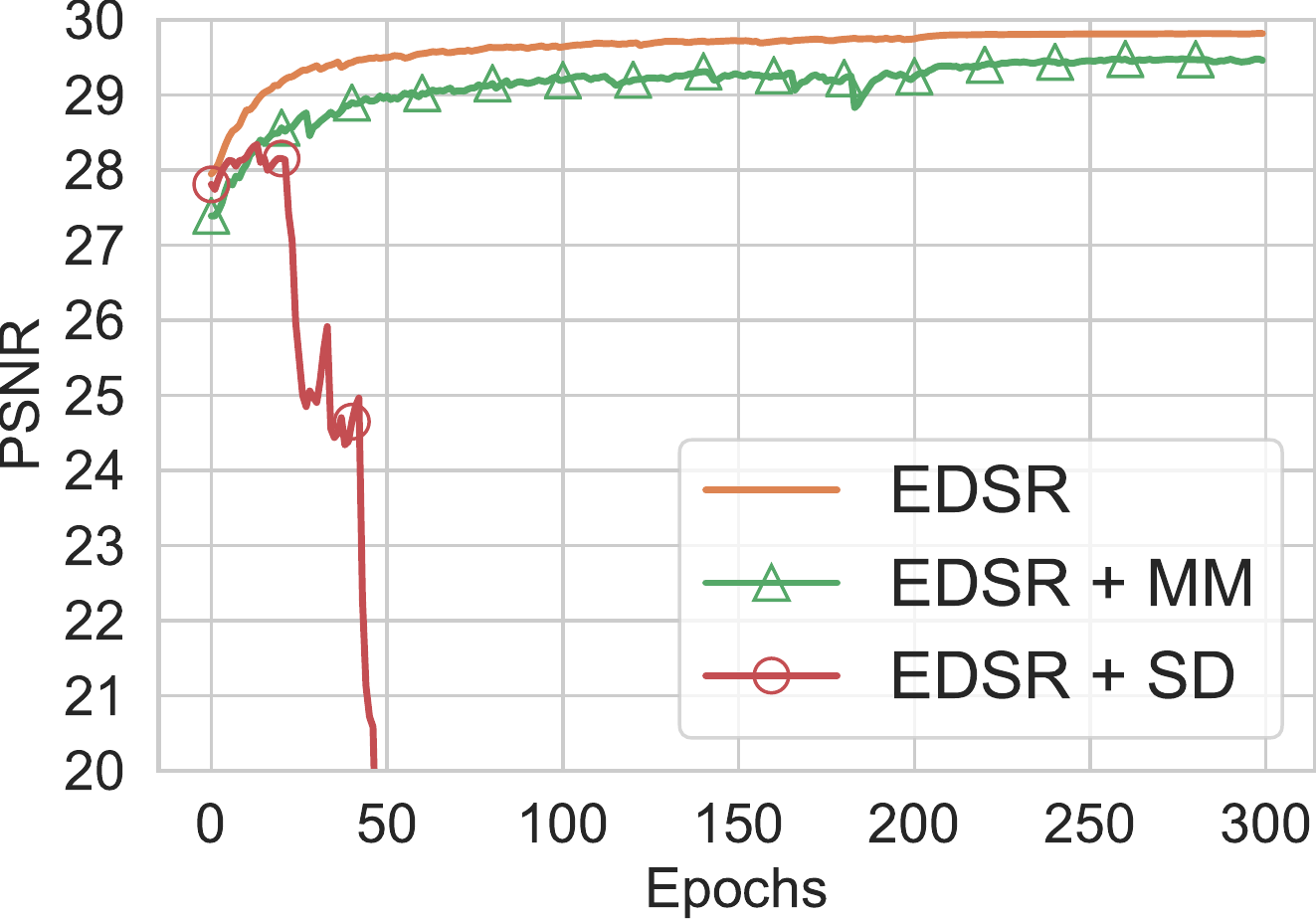}}
\caption{PSNR (dB) comparison on ten DIV2K ($\times4$) validation images during training. \textit{MM} and \textit{SD} denote the model with Manifold Mixup~\cite{manifold-mixup} and ShakeDrop~\cite{shakedrop}, respectively.}
\label{figure:curve}
\end{figure}

\noindent\textbf{Augmentation in feature space.}
We apply feature augmentations~\cite{manifold-mixup,shakedrop} to EDSR~\cite{edsr} and RCAN~\cite{rcan} (\Fref{figure:curve}). Both Manifold Mixup~\cite{manifold-mixup} and ShakeDrop~\cite{shakedrop} result in inferior performance than the baselines without any augmentation. For example, RCAN fails to learn with both Manifold Mixup and ShakeDrop. For EDSR, Manifold Mixup is the only one that can be accompanied with, but it also shows significant performance drop. The reason for the catastrophic failure of ShakeDrop is because it manipulates the training signal too much, which induces serious gradient exploding.
\smallskip

\begin{figure}[t]
\centering
\includegraphics[width=0.6\linewidth]{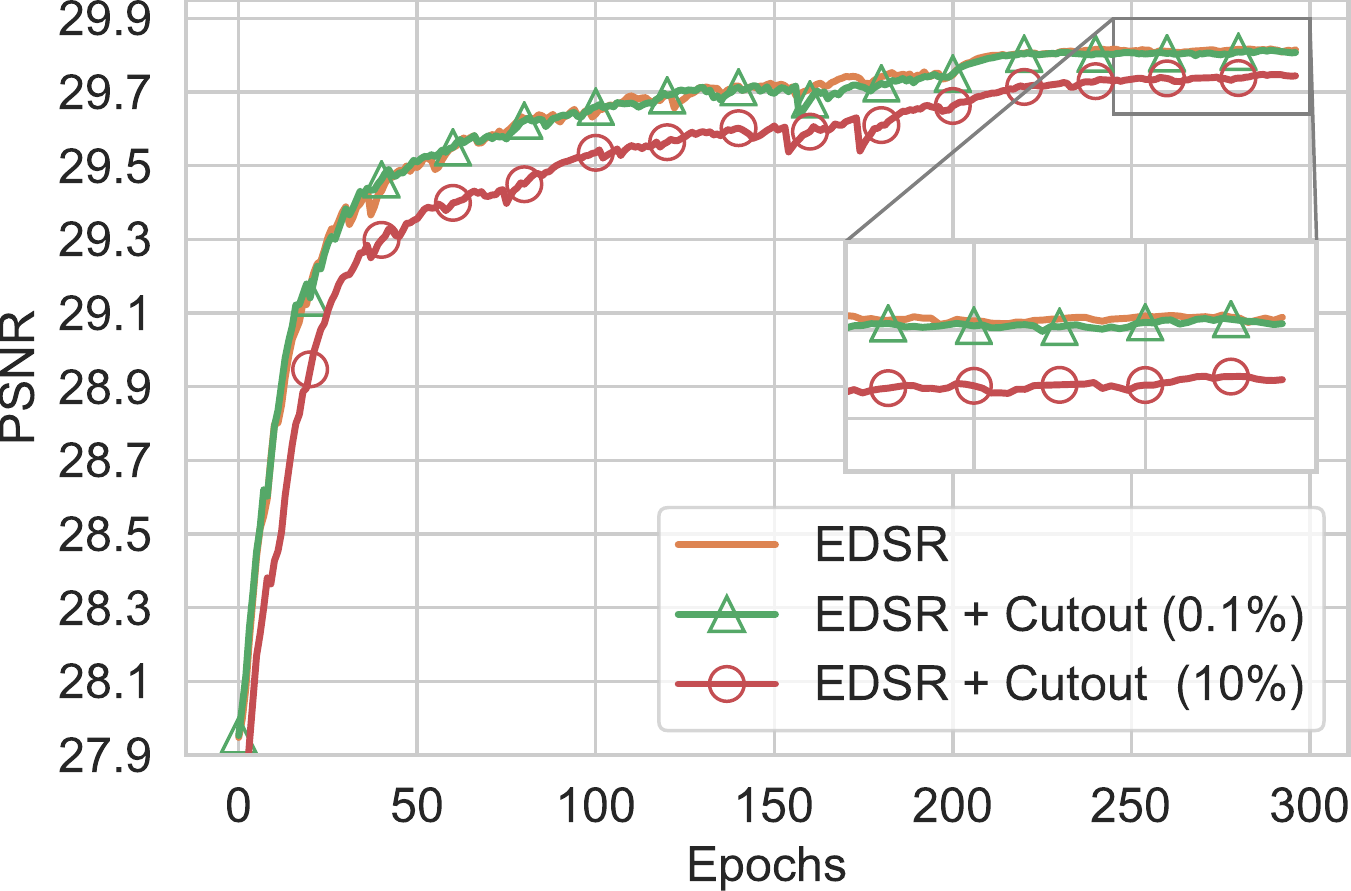}
\caption{PSNR (dB) comparison between the baseline and two Cutout~\cite{cutout} settings on ten DIV2K ($\times4$) validation images during training. The gap between two curves varies around 0.1$\sim$0.2 dB. }
\label{figure:curve-cutout}
\end{figure}

\noindent\textbf{Cutout.}
As discussed in the paper, using the original Cutout~\cite{cutout} setting seriously harms the performance. Here, we demonstrate how Cutout ratio affects the performance (\Fref{figure:curve-cutout}). Removing 0.1\% of pixels shows similar performance to the baseline, but increasing the dropping proportion to 25\% results a huge degradation.

\section{Experiment Details}
\noindent\textbf{\cutblur~vs. Giving HR inputs during training.}
For fair comparison, we provide HR images with $p=0.33$ otherwise LR images. More specifically, we set the probability of giving HR input, $p$ to 0.33, which is the same ratio to the average proportion of the HR region used in \cutblur. 
\smallskip

\noindent\textbf{Super-resolve the high resolution image.}
We quantitatively compare the performance of the baseline and \cutblur-trained model when the network takes HR images as input in the test phase (\Tref{table:hr-input}) and when the network takes \cutblur red LR input (\Tref{table:cutblur-input}). Here, we generated \cutblur red image by substituting half of the upper part of the LR to its ground truth HR image. When the network takes HR images as input, an ideal method should maintain the input resolution, which would yield infinite (dB) PSNR and 1.0 SSIM. However, the baseline (w/o \cutblur) results in a degraded performance because it tends to over-sharpen the images. This is because the model learns to blindly super-resolve every given pixel. On the other hand, our proposed method provides the near-identical image. 
when the network takes \cutblur red LR input, the performance of the models without \cutblur~are worse than the baseline (bicubic upsample kernel). In contrast, our methods achieve better performance than both the baseline (bicubic) and the models trained without \cutblur.

\begin{table}[tp]
\caption{Quantitative comparison (PSNR / SSIM / LPIPS) on the photo-realistic SR task using generative models. As a baseline, we use ESRGAN~\cite{wang2018esrgan}, which shows state-of-the-art performance on this task.}
\centering
\vspace{+0.1 cm}
\small
\begin{tabular}{c|c|c}
\hline
\multirow{2}{*}{Dataset} & ESRGAN~\cite{wang2018esrgan} & ESRGAN~\cite{wang2018esrgan} + ours \\
\cline{2-3}
& \multicolumn{2}{c}{PSNR$\uparrow$ / SSIM$\uparrow$ / LPIPS$\downarrow$} \\
\hline\hline
Set14    & 26.11 / 0.6937 / 0.143 & \textbf{26.35 / 0.7016 / 0.135} \\
B100     & 25.39 / 0.6522 / 0.177 & \textbf{25.49 / 0.6556 / 0.172} \\
Urban100 & 24.49 / 0.7364 / 0.129 & \textbf{24.54 / 0.7383 / 0.127} \\
DIV2K    & 26.52 / 0.7421 / 0.117 & \textbf{26.68 / 0.7448 / 0.114} \\
\hline
\end{tabular}
\label{table:gan}
\end{table}

\begin{table*}[ht]
\caption{Quantitative comparison (PSNR / SSIM) on artificial SR setup which gives HR image instead of LR. \textit{Baseline} indicates the quantitative metrics between the input (HR) and ground-truth (HR) images.}
\vspace{+0.1 cm}
\centering
\begin{tabular}{c|c|cc|cc}
\hline
\multirow{2}{*}{Dataset} & \multirow{2}{*}{Baseline} & \multicolumn{2}{c|}{EDSR} & \multicolumn{2}{c}{EDSR + mixture of augmentation} \\
\cline{3-6}
                         & & w/o \cutblur & w/ \cutblur & w/o \cutblur & w/ \cutblur \\
\hline\hline
DIV2K & inf. / 1.0000 & 22.61 / 0.7072 &  \textbf{\hspace{0.8em}inf. / 1.0000} & 27.33 / 0.8571 & \textbf{65.04 / 0.9999} \\
RealSR & inf. / 1.0000 & 23.23 / 0.7543 & \textbf{54.64 / 0.9985} & 24.87 / 0.8028 & \textbf{46.83 / 0.9951} \\
\hline
\end{tabular}
\label{table:hr-input}
\end{table*}

\begin{table*}[h!]
\caption{Quantitative comparison (PSNR / SSIM) on artificial SR setup which gives \cutblur red image instead of LR. We generate \cutblur red image by replacing half of the upper region of the LR to HR. \textit{Baseline} indicates the quantitative metrics between the input (\cutblur red) and ground-truth (HR) images.}
\vspace{+0.1 cm}
\centering
\begin{tabular}{c|c|cc|cc}
\hline
\multirow{2}{*}{Dataset} & \multirow{2}{*}{Baseline} & \multicolumn{2}{c|}{EDSR} & \multicolumn{2}{c}{EDSR + mixture of augmentation} \\
\cline{3-6}
                         & & w/o \cutblur & w/ \cutblur & w/o \cutblur & w/ \cutblur \\
\hline\hline
DIV2K & 29.64 / 0.8646 & 24.08 / 0.7509 & \textbf{32.09 / 0.9029} & 28.48 / 0.8403 & \textbf{32.11 / 0.9032} \\
RealSR & 30.60 / 0.8883 & 26.26 / 0.7974 & \textbf{32.50 / 0.9143} & 27.31 / 0.8244 & \textbf{32.46 / 0.9131} \\
\hline
\end{tabular}
\label{table:cutblur-input}
\end{table*}

Such observations are consistently found when using the mixture of augmentations. Note that although the model without \cutblur~(use all the augmentations except \cutblur) can improve the generalization ability compared to the vanilla EDSR, it still fails to learn such good properties of \cutblur. Only when we include \cutblur~as one of the augmentation pool, the model could learn not only ``how" but also ``where" to super-resolve an image while boosting its generalization ability in a huge margin.

\smallskip

\noindent\textbf{GAN-based SR models.}
We also apply MoA to the GAN-based SR network, ESRGAN~\cite{esrgan} and investigated the effect. 
ESRGAN is designed to produce photo-realistic SR image by adopting adversarial loss~\cite{gan}.
As shown in \Tref{table:gan}, ESRGAN with proposed method outperforms the baseline for both distortion- (PSNR and SSIM) and perceptual-based (LPIPS) metrics.
Such result implies that our method adequately enhance the GAN-based SR model as well, considering the perception-distortion trade-off~\cite{blau2018perception}.
\smallskip

\noindent\textbf{Gaussian denoising (color).}
To simulate the over-smoothing problem in the denoising task, we conduct a cross-level benchmark test (\Tref{table:cross-noise}) on various noise levels ($\sigma=[30, 50, 70]$) using  EDSR~\cite{edsr} and RDN~\cite{rdn} models.
In this setting, we test the trained networks on an unseen noise-level dataset.
We would like to emphasize that such scenario is common since we cannot guarantee that distortion information are provided in advance in real-world applications.
Here, we apply Gaussian noise to the color (RGB) image when we generate a dataset, and PSNR and SSIM are calculated on the full-RGB dimension.

\begin{figure*}[ph]
\centering
\includegraphics[width=\linewidth]{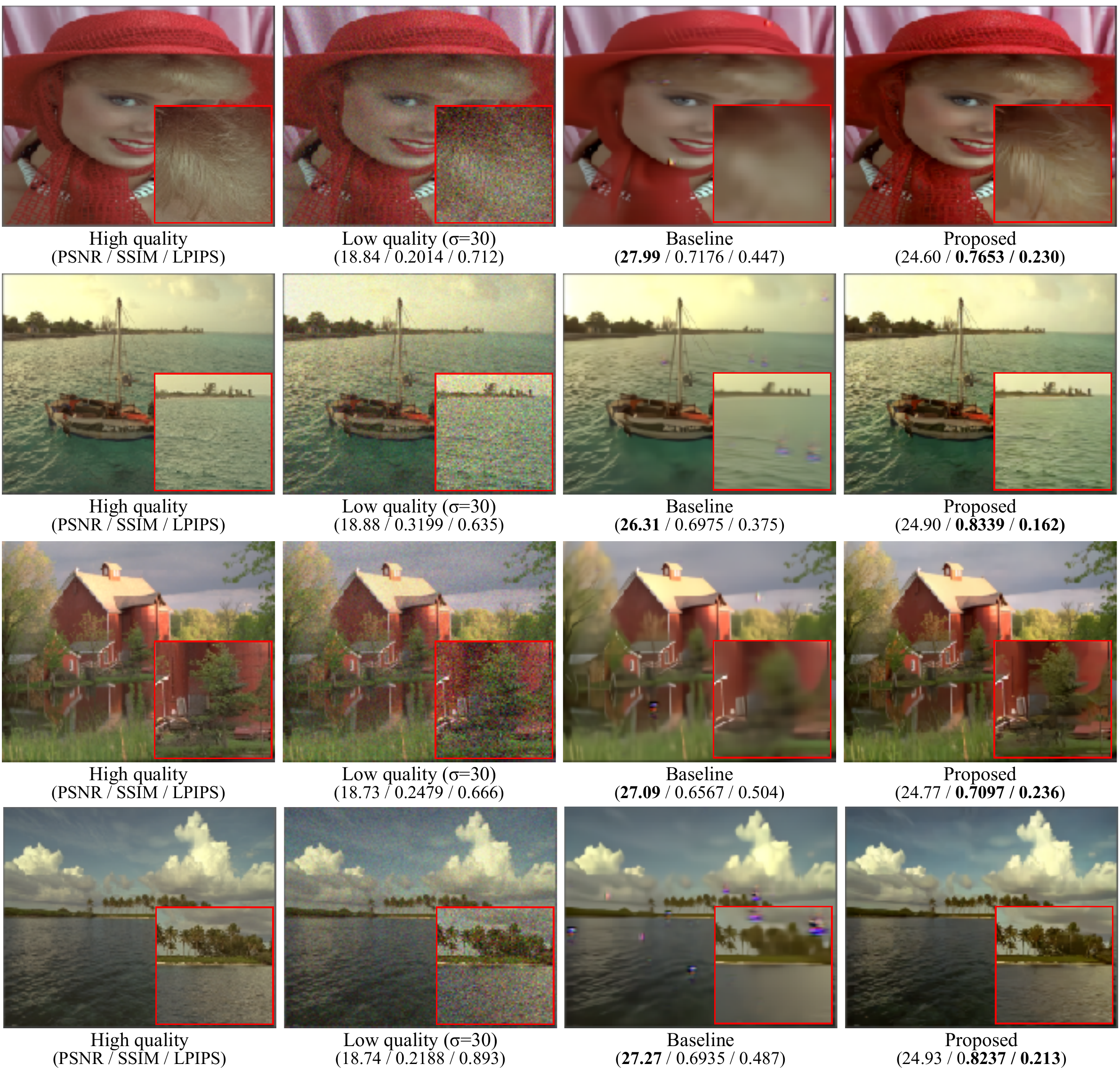}
\caption{Comparison of the generalization ability on the color Gaussian denoising task. Both methods are trained on severely distorted dataset ($\sigma=70$) and tested on the mild case ($\sigma=30$). The baseline over-smooths the inputs or generates artifacts while ours successfully reconstructs the fine structures.}
\label{fig:dn-general}
\end{figure*}

When we train the model on a mild noise level and test to a severe noise (\eg $\sigma=30\rightarrow50$), both the baseline and proposed models show degraded performance since they cannot fully eliminate a noise.
On the other hand, for severe $\rightarrow$ mild scenario, models trained with MoA surpass the baseline on SSIM and LPIPS metrics.
Note that the high PSNR scores of the baselines without MoA is due to the over-smoothing, which is preferred by PSNR. This can be easily seen in the additional qualitative results \Fref{fig:dn-general}. Interestingly, the baseline model tends to generate severe artifacts (4th row, 3rd column) since it handles unseen noise improperly. In contrast, our proposed method does not have such artifacts while effectively recovering clean images.
\smallskip

\begin{figure*}[ph]
\centering
\includegraphics[width=\linewidth]{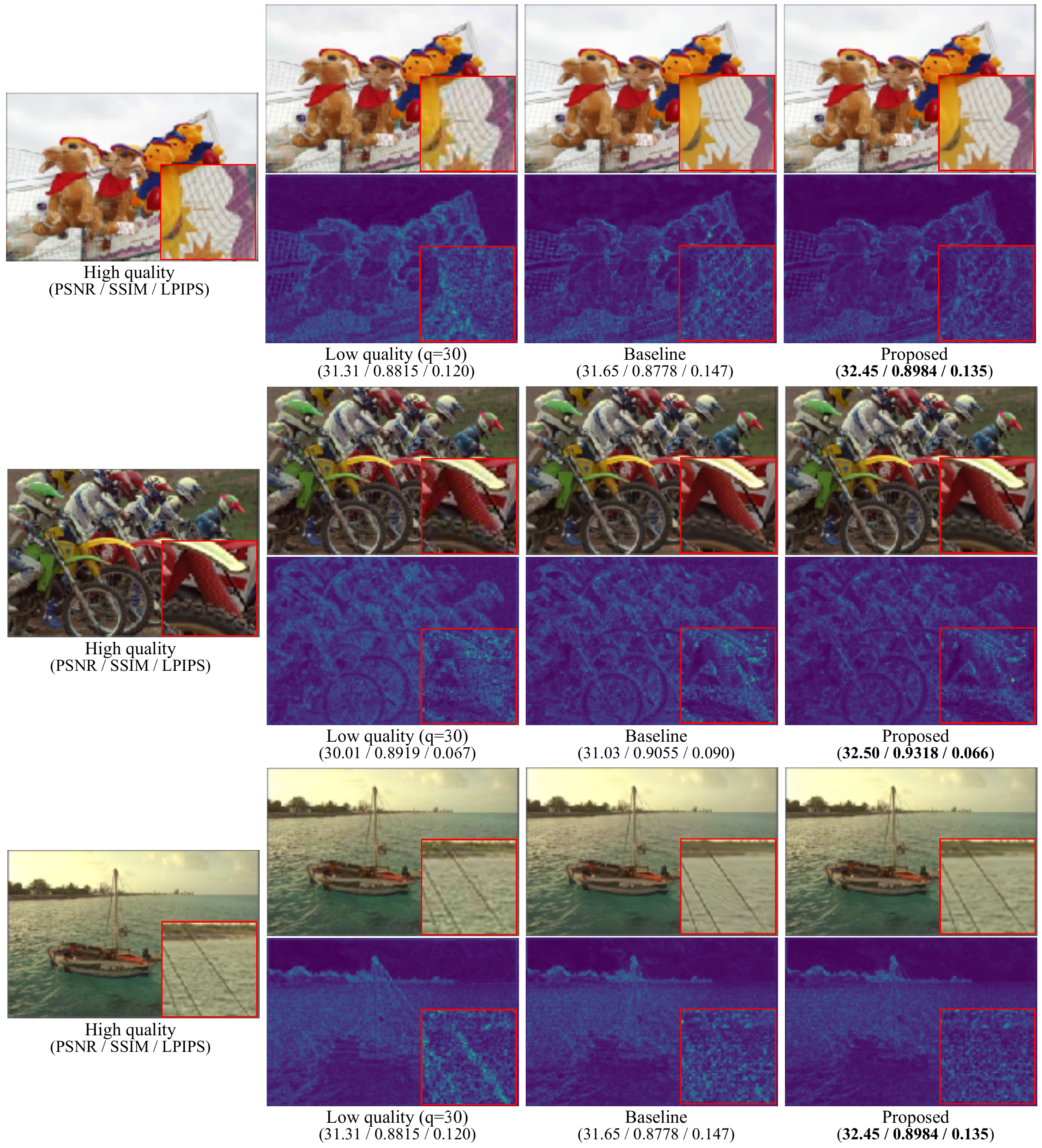}
\caption{Comparison of the generalization ability on the color JPEG artifact removal task. Both methods are trained on severely compressed dataset ($q=10$) and tested on the mild case ($q=30$).}
\label{fig:jpeg-general}
\end{figure*}

\noindent\textbf{JPEG artifact removal (color).}
Similar to the Gaussian denoising, we train and test the model with various compression factors ($q=[30, 20, 10]$).
To generate a dataset, we compress color (RGB) images with different quality levels. However, unlike the color image denoising task, we use Y channel only when calculating PSNR and SSIM. 
Quantitative and qualitative results on this task are shown in \Tref{table:cross-jpeg} and \Fref{fig:jpeg-general}, respectively.
\smallskip

\begin{figure*}[ph]
\centering
\vfill
\includegraphics[width=\linewidth]{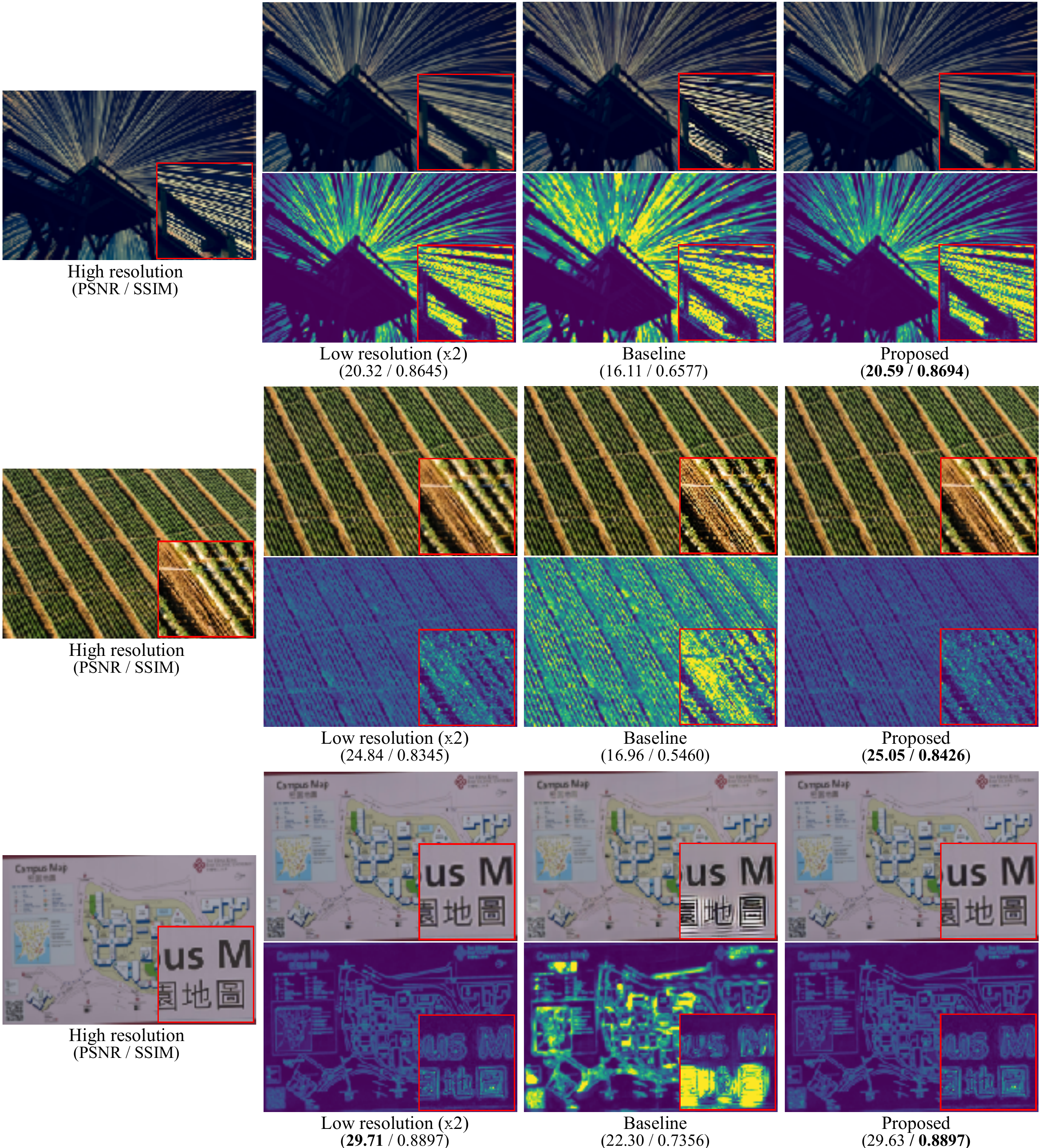}
\caption{Comparison of the generalization ability on the SR task. Both methods are trained on $\times4$ scale factor dataset and tested on different scale factor ($\times2$). The baseline tend to produce the distortion due to the over-sharpening while proposed method does not. Similar to the denoising task, the baseline over-smooths inputs so that it fails to recover fine details.}
\label{fig:sr-general}
\vfill
\end{figure*}

\noindent\textbf{Super-resolution on unseen scale factor.}
We also investigate the generalization ability of our model to the SR task. To do that, we test the models on unseen scale factors ($\times2$ and $\times3$). Here, the models are only trained on the $\times4$ scale (\Tref{table:sr-general}). Our proposed method outperforms the baseline in various scales and datasets. This tendency is more significant when the train-test mismatch becomes bigger (\eg, scale $\times2$). \Fref{fig:sr-general} shows the qualitative comparison of the baseline and ours. While the baseline model over-sharpens the edges producing embossing artifacts, our proposed method effectively super-resolve LR images of the unseen scale factor during training.
\smallskip

\begin{figure*}[ph!]
\centering
\includegraphics[width=\linewidth]{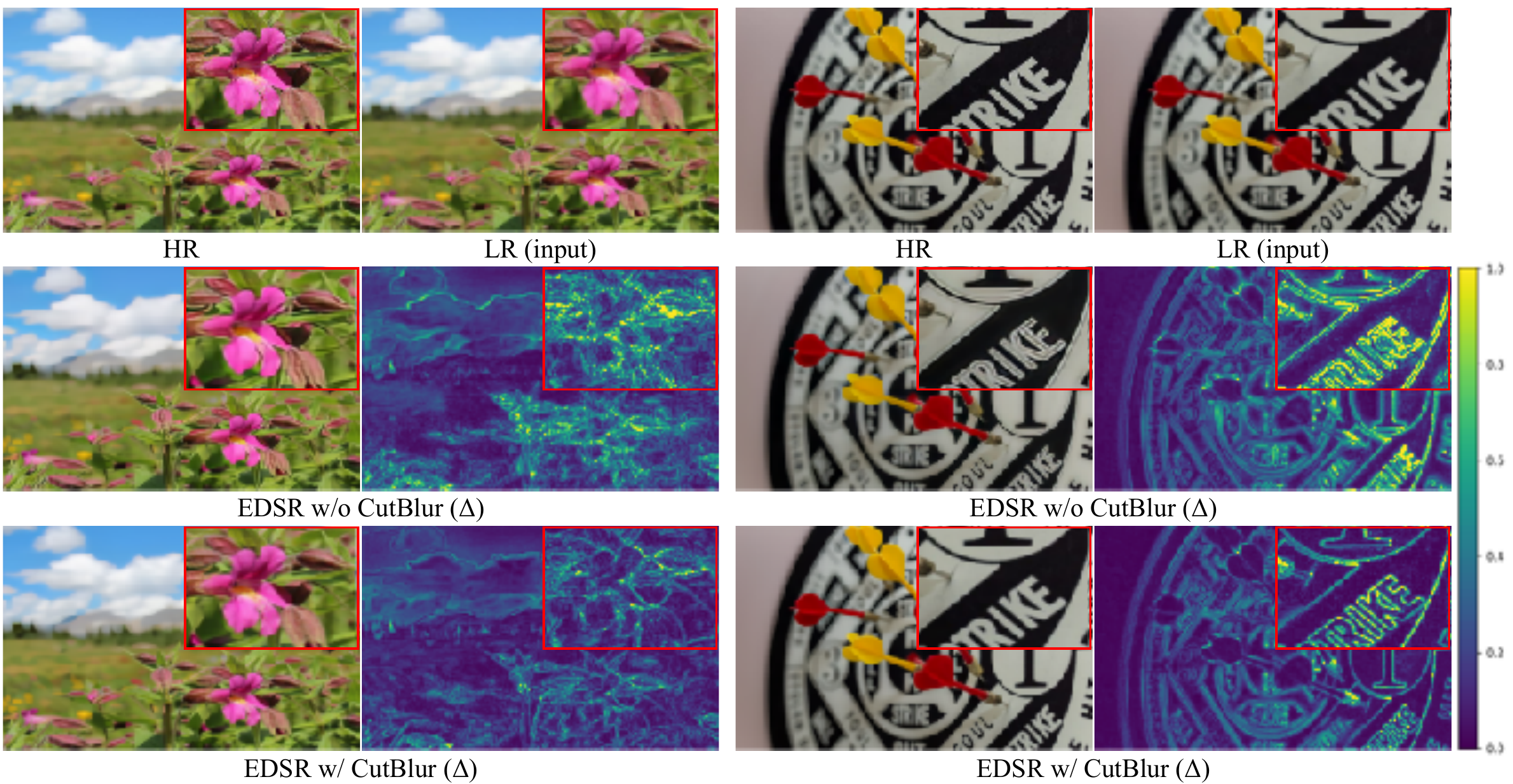}
\caption{Qualitative comparison of the baseline and \cutblur~model outputs. The inputs are the real-world out-of-focus photography ($\times2$ bicubic downsampled) taken from a web. The baseline model over-sharpens the focused region (foreground) resulting in unpleasant artifact while our method effectively super-resolves the image without generating such distortions.}
\label{fig:out-of-focus}
\end{figure*}

\noindent\textbf{\cutblur~in the wild.}
We provide more results on real-world out-of-focus photographs that are collected from web (\Fref{fig:out-of-focus}).

\begin{table*}[h!]
\caption{Performance comparison on the SR task evaluated on the DIV2K and RealSR dataset. We train the model using scale factor 4 case and test to scale factor 2 and 3.}
\vspace{+0.1 cm}
\centering
\begin{tabular}{l|c|c||c}
\hline
\multirow{2}{*}{Model} & \multirow{2}{*}{Test Scale} & \multicolumn{2}{c}{Train Scale ($\times4$)} \\
\cline{3-4}
& & DIV2K & RealSR \\
\hline\hline
EDSR     & \multirow{2}{*}{$\times2$} & 23.75 (+0.00) / 0.7414 (+0.0000) & 27.51 (+0.00) / 0.8273 (+0.0000) \\
+ proposed     &                      & 31.27 \textbf{\green{(+7.52)}} / 0.8970 \textbf{\green{(+0.1556)}} & 31.61 \textbf{\green{(+4.10)}} / 0.8985 \textbf{\green{(+0.0712)}} \\
\hline
EDSR     & \multirow{2}{*}{$\times3$} & 27.62 (+0.00) / 0.8142 (+0.0000) & 29.44 (+0.00) / 0.8467 (+0.0000) \\
+ proposed     &                      & 28.40 \textbf{\green{(+0.78)}} / 0.8170 \textbf{\green{(+0.0028)}} & 29.94 \textbf{\green{(+0.50)}} / 0.8542 \textbf{\green{(+0.0075)}} \\
\hline
\end{tabular}
\label{table:sr-general}
\end{table*}

\begin{table*}[t!]
\centering
\caption{Performance comparison on the color Gaussian denoising task evaluated on the Kodak24 dataset. We train and test the model on the various noise levels. LPIPS \cite{lpips} (lower is better) indicates the perceptual distance between the network output and the ground-truth.}
\begin{tabular}{c|c|ccc}
\hline
\multirow{2}{*}{Model} & \multirow{2}{*}{Train $\sigma$} & Test ($\sigma=30$) & Test ($\sigma=50$) & Test ($\sigma=70$) \\
\cline{3-5}
& & PSNR$\uparrow$ / SSIM$\uparrow$ / LPIPS$\downarrow$ & PSNR$\uparrow$ / SSIM$\uparrow$ / LPIPS$\downarrow$ & PSNR$\uparrow$ / SSIM$\uparrow$ / LPIPS$\downarrow$ \\
\hline\hline
EDSR     & \multirow{4}{*}{30} & 31.92 / \hspace{1ex}0.8716 / \hspace{1ex}0.136 & 
                                 20.78 / \hspace{1ex}0.3425 / \hspace{1ex}0.690 & 
                                 16.38 / \hspace{1ex}0.1867 / \hspace{1ex}1.004 \\
+ proposed & & \bgreen{+0.02} / \bgreen{+0.0006} / \bgreen{\minus0.004} & \bgreen{+1.05} / \bgreen{+0.0446} / \bgreen{\minus0.105} & \bgreen{+0.35} / \bgreen{+0.0145} / \bgreen{\minus0.062}\\
RDN     &  &                     31.92 / \hspace{1ex}0.8715 / \hspace{1ex}0.137 & 
                                 21.61 / \hspace{1ex}0.3733 / \hspace{1ex}0.639 & 
                                 16.99 / \hspace{1ex}0.2040 / \hspace{1ex}0.974 \\
+ proposed & & +0.00 / \bgreen{+0.0002} / +0.000 & \red{\minus1.01} / \red{\minus0.0368} / \red{+0.016} & \red{\minus0.61} / \red{\minus0.0182} / \bgreen{\minus0.020} \\
\hline
EDSR     & \multirow{4}{*}{50} & 29.64 / \hspace{1ex}0.7861 / \hspace{1ex}0.306 & 
                                 29.66 / \hspace{1ex}0.8136 / \hspace{1ex}0.209 & 
                                 21.50 / \hspace{1ex}0.3553 / \hspace{1ex}0.687 \\
+ proposed & & \red{\minus0.54} / \bgreen{+0.0708} / \bgreen{\minus0.158} &                          +0.00 / \red{\minus0.0002} / \bgreen{\minus0.001} &
               \bgreen{+0.26} / \bgreen{+0.0212} / \bgreen{\minus0.029}\\
RDN     &                  &     29.77 / \hspace{1ex}0.7931 / \hspace{1ex}0.298 & 
                                 29.63 / \hspace{1ex}0.8134 / \hspace{1ex}0.208 & 
                                 23.68 / \hspace{1ex}0.4549 / \hspace{1ex}0.519 \\
+ proposed & & \red{\minus1.00} / \bgreen{+0.0544} / \bgreen{\minus0.146} &                          \red{\minus0.01} / \red{\minus0.0005} / \red{+0.002} &
               \red{\minus1.40} / \red{\minus0.0666} / \red{+0.104}\\
\hline

EDSR     & \multirow{4}{*}{70} & 27.38 / \hspace{1ex}0.7295 / \hspace{1ex}0.375 & 
                                 27.95 / \hspace{1ex}0.7385 / \hspace{1ex}0.366 & 
                                 28.23 / \hspace{1ex}0.7689 / \hspace{1ex}0.273 \\
+ proposed & & \red{\minus2.51} / \bgreen{+0.0696} / \bgreen{\minus0.193} &                \bgreen{+0.46} / \bgreen{+0.0674} / \bgreen{\minus0.139} &
               +0.00 / \red{\minus0.0003} / \bgreen{\minus0.002}\\
RDN     &                      & 28.23 / \hspace{1ex}0.7546 / \hspace{1ex}0.344 & 
                                 28.13 / \hspace{1ex}0.7517 / \hspace{1ex}0.349 & 
                                 28.19 / \hspace{1ex}0.7684 / \hspace{1ex}0.275 \\
+ proposed & & \red{\minus3.48} / \bgreen{+0.0461} / \bgreen{\minus0.163} &                \red{\minus0.93} / \bgreen{+0.0337} / \bgreen{\minus0.137} &
               \bgreen{+0.01} / \red{\minus0.0003} / \bgreen{\minus0.006}\\
\hline
\end{tabular}
\label{table:cross-noise}
\end{table*}

\begin{table*}[t!]
\centering
\caption{Performance comparison on the color JPEG artifact removal task evaluated on the LIVE1~\cite{sheikh2005live} dataset. We train and test the model on the various quality factors. LPIPS \cite{lpips} (lower is better) indicates the perceptual distance between the network output and the ground-truth. Unlike the color image denoising task, we use Y channel only when calculating PSNR and SSIM.}
\begin{tabular}{c|c|ccc}
\hline
\multirow{2}{*}{Model} & \multirow{2}{*}{Train $q$} & Test ($q=30$) & Test ($q=20$) & Test ($q=10$) \\
\cline{3-5}
& & PSNR$\uparrow$ / SSIM$\uparrow$ / LPIPS$\downarrow$ & PSNR$\uparrow$ / SSIM$\uparrow$ / LPIPS$\downarrow$ & PSNR$\uparrow$ / SSIM$\uparrow$ / LPIPS$\downarrow$ \\
\hline\hline
EDSR     & \multirow{4}{*}{30} & 33.95 / \hspace{1ex}0.9227 / \hspace{1ex}0.118 & 
                                 32.36 / \hspace{1ex}0.8974 / \hspace{1ex}0.155 & 
                                 29.17 / \hspace{1ex}0.8196 / \hspace{1ex}0.286 \\
+ proposed & & \red{\minus0.01} / \red{\minus0.0002} / \red{+0.001} & \bgreen{+0.02} / \bgreen{+0.0003} / \red{+0.001} & +0.00 / \red{\minus0.0005} / \red{+0.001}\\
RDN     &  &                     33.90 / \hspace{1ex}0.9220 / \hspace{1ex}0.121 & 
                                 32.34 / \hspace{1ex}0.8971 / \hspace{1ex}0.157 & 
                                 29.20 / \hspace{1ex}0.8202 / \hspace{1ex}0.287 \\
+ proposed & & \bgreen{+0.01} / \bgreen{+0.0003} / \bgreen{\minus0.003}  & \red{\minus0.01} / \red{\minus0.0001} / \bgreen{\minus0.001} & \red{\minus0.05} / \red{\minus0.0017} / \red{+0.002} \\
\hline
EDSR     & \multirow{4}{*}{20} & 33.64 / \hspace{1ex}0.9174 / \hspace{1ex}0.130 & 
                                 32.52 / \hspace{1ex}0.8979 / \hspace{1ex}0.160 & 
                                 29.67 / \hspace{1ex}0.8327 / \hspace{1ex}0.271 \\
+ proposed & & \bgreen{+0.18} / \bgreen{+0.0037} / \bgreen{\minus0.008} &                          +0.00 / \bgreen{+0.0001} / \bgreen{\minus0.001} &
               +0.00 / \red{\minus0.0005} / \bgreen{\minus0.001}\\
RDN     &                  &     33.59 / \hspace{1ex}0.9164 / \hspace{1ex}0.132 & 
                                 32.47 / \hspace{1ex}0.8972 / \hspace{1ex}0.162 & 
                                 29.65 / \hspace{1ex}0.8322 / \hspace{1ex}0.271 \\
+ proposed & & \bgreen{+0.14} / \bgreen{+0.0031} / \bgreen{\minus0.005} &                          +0.00 / \bgreen{+0.0001} / \red{+0.001} &
               \bgreen{+0.01} / \red{\minus0.0003} / \red{+0.001}\\
\hline

EDSR     & \multirow{4}{*}{10} & 32.45 / \hspace{1ex}0.8992 / \hspace{1ex}0.154 & 
                                 31.83 / \hspace{1ex}0.8840 / \hspace{1ex}0.179 & 
                                 30.14 / \hspace{1ex}0.8391 / \hspace{1ex}0.254 \\
+ proposed & & \bgreen{+0.97} / \bgreen{+0.0179} / \bgreen{\minus0.020} &                \bgreen{+0.45} / \bgreen{+0.0104} / \bgreen{\minus0.011} &
               +0.00 / \red{\minus0.0001} / \red{+0.001}\\
RDN     &                      & 32.37 / \hspace{1ex}0.8967 / \hspace{1ex}0.166 & 
                                 31.78 / \hspace{1ex}0.8821 / \hspace{1ex}0.189 & 
                                 30.10 / \hspace{1ex}0.8381 / \hspace{1ex}0.259 \\
+ proposed & & \bgreen{+0.95} / \bgreen{+0.0187} / \bgreen{\minus0.023} &                \bgreen{+0.40} / \bgreen{+0.0106} / \bgreen{\minus0.013} &
               \red{\minus0.01} / \red{\minus0.0002} / \red{+0.003}\\
\hline
\end{tabular}
\label{table:cross-jpeg}
\end{table*}

\vspace{10 cm}

\clearpage
\balance

{\small
\bibliographystyle{ieee_fullname}
\bibliography{}
}

\end{document}